\documentclass[%
reprint,
showkeys,
nofootinbib,
amsmath,amssymb,
aps,
prl,
]{revtex4-1}
\usepackage[marginal]{footmisc}
\usepackage{graphicx}
\usepackage{dcolumn}
\usepackage{bm}
\usepackage[colorlinks,allcolors=black,CJKbookmarks=True]{hyperref}

\begin{document}
	
	\title{Primordial black hole production during first-order phase transitions}
	
	\author{Jing Liu$^{1,2}$}
	\email{liujing@ucas.ac.cn}
	
	\author{Ligong Bian$^{3,4}$}
	\email{lgbycl@cqu.edu.cn}
	
	\author{Rong-Gen Cai$^{5,2,1}$}
	\email{cairg@itp.ac.cn}
	
	\author{Zong-Kuan Guo$^{5,2,1}$}
	\email{guozk@itp.ac.cn}
	
	\author{Shao-Jiang Wang$^{5}$}
	\email{schwang@itp.ac.cn}
	
	\affiliation{$^{1}$School of Fundamental Physics and Mathematical Sciences, Hangzhou Institute for Advanced Study, University of Chinese Academy of Sciences, Hangzhou 310024, China
	}
	
	\affiliation{$^{2}$School of Physical Sciences, University of Chinese Academy of Sciences,
		No.19A Yuquan Road, Beijing 100049, China}
	
	\affiliation{$^{3}$Department of Physics, Chongqing University, Chongqing 401331, China}
	
	\affiliation{$^{4}$Chongqing Key Laboratory for Strongly Coupled Physics, Chongqing 401331, China}
	
	\affiliation{$^{5}$CAS Key Laboratory of Theoretical Physics, Institute of Theoretical Physics,
		Chinese Academy of Sciences, P.O. Box 2735, Beijing 100190, China}

	\begin{abstract}
		A novel mechanism for the primordial black hole (PBH) production is proposed as a natural and inevitable consequence of general first-order phase transitions without reference to specific underlying particle physics models.
		We obtain mutual predictions and constraints between primordial black holes and gravitational waves from phase transitions in the general case. 
		For particular interest, our PBHs generated during a PeV-scale phase transition could make up all the dark matter, while PBHs generated during a MeV-scale phase transition could simultaneously account for LIGO-Virgo coalescence events and NANOGrav 12.5-yr result for the corresponding gravitational waves. 
		
	\end{abstract}
	
	\maketitle

	\emph{Introduction}. 
	The detection of gravitational waves~(GWs) from the early Universe opens a new window to probe new physics~\cite{Cai:2017cbj,Bian:2021ini}. 
	Among them, the first-order PTs are well-motivated and predicted by many extensions of the Standard Model of particle physics~(see, for example, \cite{Caprini:2019egz} for a recent summary.)
	During a PT, the vacuum bubbles copiously nucleate due to quantum tunneling~\cite{Coleman:1977py,Callan:1977pt} and continuously release the vacuum energy into bubble walls and background plasma~\cite{Espinosa:2010hh,Ellis:2019oqb,Cai:2020djd}. As the vacuum pressure pushes bubble walls forward, true vacuum bubbles continuously expand and finally collide with each other, generating large energy density perturbations and then produce GWs.

	Furthermore, it has been proposed that the collapse of false vacuum bubbles formed during bubble collisions with a certain bubble-formation rate leads to the formation of primordial black holes (PBHs)~\cite{Hawking:1982ga,Crawford:1982yz,Kodama:1982sf,Johnson:2011wt,Khlopov:1998nm}. PBHs are of broad interest for the ability to constitute even all dark matter~\cite{Carr:2016drx,Carr:2020xqk} and explain the coalescence events observed by LIGO-Virgo collabration~\cite{Sasaki:2016jop,Bird:2016dcv}.
	Generally speaking, amplified energy density perturbations also provide the necessary conditions of PBH formation when generating GWs. Once energy density fluctuations reach the threshold, PBHs formed from the gravitational collapse of the whole mass within a Hubble horizon~\cite{Niemeyer:1997mt,Kuhnel:2015vtw}. In this letter, we propose a novel and generic mechanism of PBH formation during PTs arising from the asynochronization of the vacuum tunneling progress, which is different from the PBH formations mechanism proposed in Ref.~\cite{Gross:2021qgx,Baker:2021nyl,Kawana:2021tde} that proceeding through the collapse of fermion bubbles depending on the interaction between the scalar field and background plasma~\cite{Gross:2021qgx,Baker:2021nyl,Kawana:2021tde}. 
	
	In this letter, we do not assume any specific  particle physics model for PT to achieve such a PBH production channel, and find the PBH formation is a universal consequence of PTs. Because of the asynochronization of the vacuum tunneling progress, there always exist some Hubble-sized regions where the decay of the false vacuum is postponed. Since the radiation energy density decreases in the expanding Universe while the vacuum energy remains constant, the postpone of false vacuum decay leads to an increase in total energy density within those regions. At the time the vacuum energy inside those regions totally decays into other components, the overdensity reaches the maxmum and the whole mass inside those regions may collapse into PBHs. Different from the previous works, we utilize the concrete numerical result of gravitational collapse~\cite{Musco:2004ak,Harada:2013epa} to obtain the exact result of the PBH abundance in terms of the basic model parameters of PTs, where numerical relativity simulation results are indispensable to obtain correct PBH formation mechanism. Especially PBHs abundantly form in a relatively slow and supercooled PT.
	We also discuss the mutual constraints on GW energy spectrum and PBH abundance basing on our mechanism. 
	For convenience, we choose $c=8\pi G=1$ throughout this letter.

	\emph{PBHs from PTs}.
	The exponential nucleation rate of vacuum bubbles reads~\cite{Coleman:1977py,Enqvist:1991xw}
	\begin{equation}
		\label{eq:Gamma}
		\Gamma(t)=\Gamma_{0}e^{\beta t}\,,
	\end{equation}
	where $\Gamma_{0}$ is the initial value, the key parameter $\beta\geq 0$ represents the increaing rate. For the case $\beta\gg 1$, $\beta^{-1}$ is also an estimation of the duration time of PTs.
	The averaged spatial fraction of the false vacuum reads~\cite{Turner:1992tz}
	\begin{equation}
		\label{eq:Ftn}
		F(t)=\exp \left[-\frac{4 \pi}{3} \int_{t_{i}}^{t} \mathrm{~d} t^{\prime} \Gamma\left(t^{\prime}\right) a^{3}\left(t^{\prime}\right) r^{3}\left(t, t^{\prime}\right)\right]\,,
	\end{equation}
	where $r(t,t')\equiv\int_{t'}^{t} a^{-1}(\tau) d\tau$ denotes the comoving radius of the true vacuum bubble and $t_{i}$ is the nuleation time of the first bubble. Before $t_{i}$ the false vacuum covers the whole space so $F(t)=1$, and then $F(t)$ gradually decreases with the expansion of true vacuum bubbles. As $F(t)$ decreases, the vacuum energy is released to bubble walls and background plasma.
	The Friedmann equation reads
	\begin{equation}\label{eq:fried}
		H^{2}=\frac{\rho_{\mathrm{v}}+\rho_{\mathrm{r}}+\rho_{\mathrm{w}}}{3}\,,
	\end{equation}
	where $\rho_{\mathrm{v}}$, $\rho_{\mathrm{r}}$ and $\rho_{\mathrm{w}}$ are the energy densities of false vacuum, background radiation and bubble walls, respectively. Here $\rho_{\mathrm{v}}=F(t)\Delta V$ and $\Delta V$ is the energy density difference between the true and false vacua, where we have normalized the true vacuum energy density to be the zero-point of the vacuum energy.
	$\alpha(t)\equiv \Delta V/\rho_{\mathrm{r}}(t)$ is another key parameter which represents the the ratio of the vacuum energy density released to the radiation bath during PTs. The initial value of $\alpha(t)$ determines the strength of the PT which is given later in specific cases.
	We assume the velocity of bubble walls is close to the speed of light. Thus, the equation of state parameters of both bubble walls and background plasma are $1/3$.
	Then, the evolution of radiation energy density reads 
	\begin{equation}\label{eq:rho}
		\frac{d(\rho_{\mathrm{r}}+\rho_{\mathrm{w}})}{dt}+4H(\rho_{\mathrm{r}}+\rho_{\mathrm{w}})=\left(-\frac{d\rho_{\mathrm{v}}}{dt}\right)\,,
	\end{equation}
	where the l.h.s. represents the Hubble expansion effect and the r.h.s. comes from the release of vacuum energy.
	Since the vacuum decay is probabilistic, there always exists some probability that in some Hubble volumes the vacuum decay is postponed, the corresponding probability is obtained from Eq.~\eqref{eq:Gamma},
	\begin{equation}
		\label{eq:prob}
		P(t_{n})=\exp\left[-\frac{4\pi}{3}\int_{t_{i}}^{t_{n}}\frac{a^{3}(t)}{a^{3}(t_{\mathrm{PBH}})}H^{-3}(t_{\mathrm{PBH}})\Gamma(t) dt\right]\,,
	\end{equation}
	where $t_{n}$ is the time for bubbles starting to nucleate in those Hubble volumes with postponed vacuum decay and $t_{\mathrm{PBH}}$ is the PBH formation time.
	
	 In the expanding Universe $\rho_{\mathrm{v}}$ is constant but $\rho_{\mathrm{w}},\rho_{\mathrm{r}}\propto a^{-4}(t)$, so that the delay of vacuum decay results in an increase of the total energy density.
	 The energy density ratio inside and outside those regions, $\rho_{\mathrm{inside}}/\rho_{\mathrm{outside}}$, continues increasing until vacuum energy totally decay inside the overdense regions. At that time, $\rho_{\mathrm{inside}}/\rho_{\mathrm{outside}}$ reaches the maximum and remains a constant afterwards. Then, the mixture of both radiation and relativistic bubble walls collapse into PBHs if the overdensity reaches the threshold $1+\delta_{c}$.
	 Increasing $t_{n}$ results in larger $\rho_{\mathrm{inside}}/\rho_{\mathrm{outside}}$ so that PBHs can form more easily. However, as implied by Eq.~\eqref{eq:prob}, the probability of the overdense regions quickly decreases with larger $t_{n}$, and the PBH abundance becomes smaller. Eq.~\eqref{eq:prob} also implies that $P(t_{n})$ exponentially decreases with $\beta$. In other words, although PBHs naturally form during PTs, they can constitute a nonnegligible part of dark matter only for relatively slow PTs.
	
	If the energy density in one Hubble volume exceeds the critical density for a certain threshold, $\delta_{c}=0.45$~\cite{Musco:2004ak,Harada:2013epa}, almost the whole mass in the Hubble horizon will collapse into a PBH, and the PBH mass reads
	\begin{equation}
		\label{eq:PBHm}
		M_{\mathrm{PBH}}\approx \frac{4\pi}{3}\gamma H^{-3}(t_{\mathrm{PBH}})\rho_{c}=4\pi\gamma H^{-1}(t_{\mathrm{PBH}})\,,
	\end{equation}
	where $\rho_{c}$ is the critical density and $\gamma\lesssim 1$ is a factor depending on detailed dynamics of gravitational collapse. See Refs.~\cite{Musco:2020jjb} for more accurate estimation of the PBH mass, we simply adopt Eq.~\eqref{eq:PBHm} as an illustration. 
	In the next section, we will show the evolution of $\rho_{\mathrm{v}}(t)$ and $\rho_{\mathrm{r}}(t)+\rho_{\mathrm{w}}(t)$, the amplification of energy density perturbations, as well as the predicted PBH mass function in detail.
	
	The previous analysis implies that the PBH abundance only depends on the basic parameters $\alpha$ and $\beta$, and we do not assume any details of specific models. Thus, the mechanism and result of this work are generic.
	
	In general, The PTs happen in the radiation-dominated era, and the scale factor $a$ increases for many orders of magnitude until the time of matter-radiation equality. Since the PBHs energy densities  $\rho_{\mathrm{PBH}}\propto a^{-3}$ and radiation energy density $\rho_{\mathrm{r}}\propto a^{-4}$, the proportion of PBH energy density is immensely amplified by the expansion of the Universe. 
	Thus, the PBH abundance is nonnegligible although the probability $P(t_{n})$ may be many orders of magnitude smaller than $1$.
	
	
	\begin{figure*}
		\flushleft\includegraphics[height=1.9in]{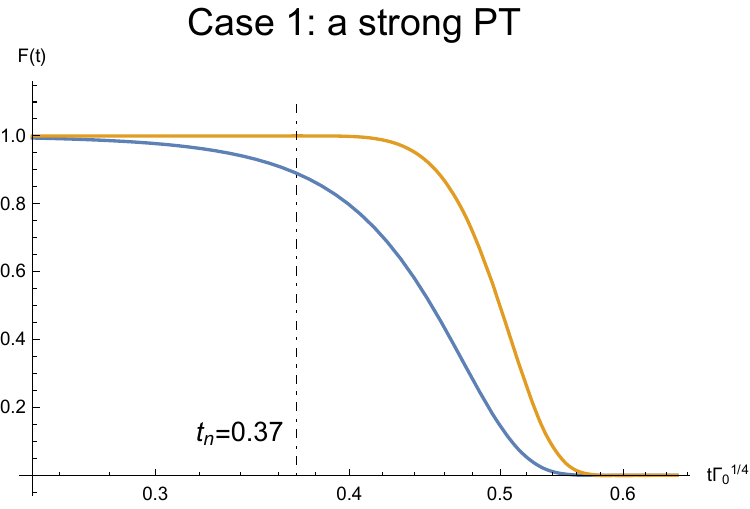}
		\includegraphics[height=1.9in]{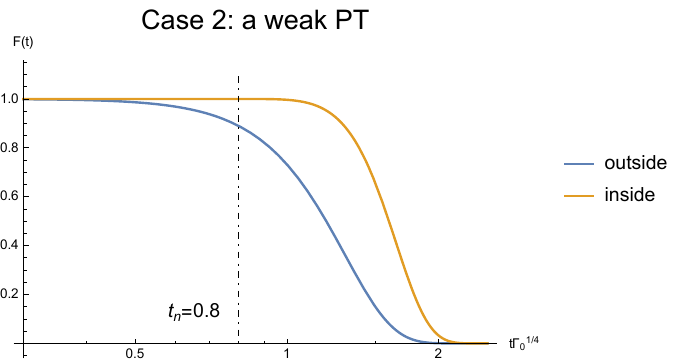}
		\flushleft\includegraphics[height=1.7in]{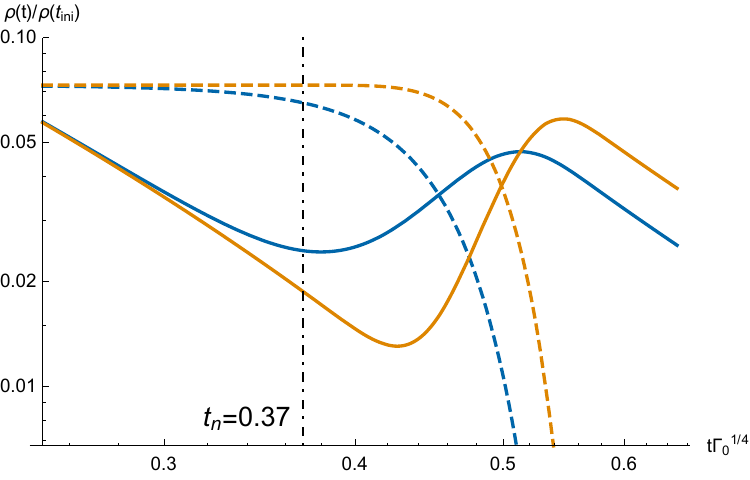}
		\includegraphics[height=1.7in]{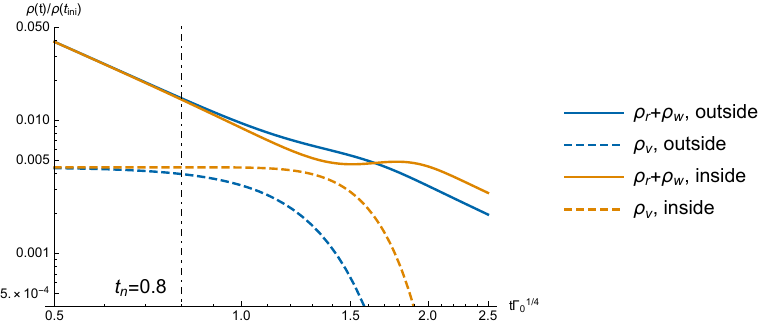}
		\caption{The evolution of the false vacuum fraction $F(t)$~(upper) and each component of the energy density~(lower) in case 1~(left) and case 2~(right). The energy density of each component is rescaled by the initial value $\rho(t_{i})$. The bule and orange lines denotes the conditions inside and outside the overdense regions. The dot-dashed line denotes the time of the first bubble nucleation $t_{n}$ inside the overdense region. The solid and dashed lines in the lower panel respectively depict the evolution of the energy density of the false vacuum and the radiation.}
		\label{fig:Frho}
	\end{figure*}
	
	\begin{figure*}
		\flushleft\includegraphics[height=1.9in]{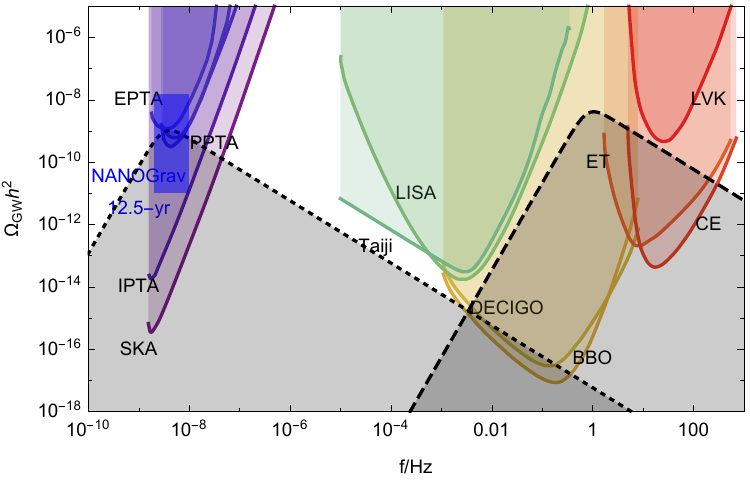}
		\includegraphics[height=2.15in]{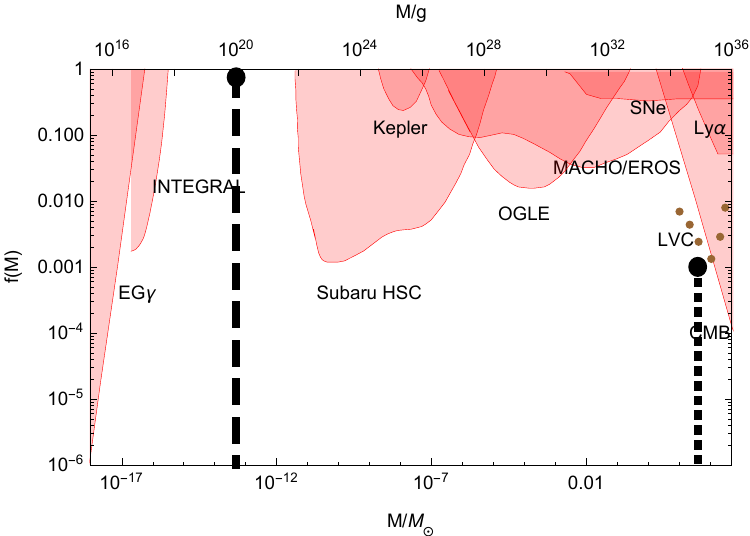}
		\raisebox{2\height}{\includegraphics[width=0.8in]{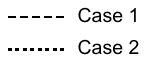}}
		\caption{The predicted GW energy spectra and PBH mass functions in each case. The sensitivity curves can be found in Ref.~\cite{Schmitz:2020syl}, including EPTA~\cite{Lentati:2015qwp}, PPTA~\cite{Shannon:2015ect}, NANOGrav~\cite{Arzoumanian:2018saf}, IPTA~\cite{Hobbs:2009yy}, SKA~\cite{Carilli:2004nx}, LISA~\cite{Audley:2017drz} Taiji~\cite{Guo:2018npi}, DECIGO~\cite{Kawamura:2011zz}, BBO~\cite{phinney2004big}, LIGO, Virgo and KAGRA~( LVK)~\cite{TheLIGOScientific:2014jea,Somiya:2011np}, CE~\cite{Reitze:2019iox}, ET~\cite{Punturo:2010zz}. The PBH constraints include EG$\gamma$~\cite{Carr:2009jm}, INTEGRAL~\cite{DeRocco:2019fjq,Laha:2019ssq,Dasgupta:2019cae}, Subaru HSC~\cite{Niikura:2017zjd}, Kepler~\cite{Griest:2013esa}, OGLE~\cite{Niikura:2019kqi}, MACHO/EROS~\cite{Allsman:2000kg,Tisserand:2006zx}, SNe~\cite{Zumalacarregui:2017qqd}, Ly$\alpha$~\cite{Murgia:2019duy}, CMB~\cite{Poulin:2017bwe} and LIGO-Virgo collabration~(LVC)~\cite{Vaskonen:2019jpv}.} 
		\label{fig:pbhgw}
	\end{figure*}
	\emph{The PBH mass function and GW energy spectrum}.
	We consider two typical cases as examples 1) a strong PT with $\alpha>1$, 2) a weak PT with $\alpha<1$, and give the evolution of each component of the energy density, then obtain the predicted GW energy spectrum and PBH abundance. 
	
	
	
	
	For the parameters we consider in the following, bubble collisions dominate the GW sources, and we neglect GWs from sound waves and magneto-hydrodynamic turbulence.
	The numerical simulations under envelope approximation give the energy spectrum of GWs from bubble collisions~\footnote{The collapse of PBHs may result in additional GW production. Since the overdense regions are very rare in the whole Universe, this SGWB is too weak compared to that sourced from bubble collisions elsewhere.}~\cite{Kamionkowski:1993fg,Huber:2008hg}
	\begin{equation}
		\begin{split}
			\Omega_{\mathrm{GW}}h^{2}(f)=1.67 \times& 10^{-5}\left(H_{*}/\beta\right)^{2}\left(\frac{\kappa \alpha_{*}}{1+\alpha_{*}}\right)^{2}\times\\
			&\frac{0.11 v_{w}^{3}}{0.42+v_{w}^{2}} \frac{3.8(f/f_{p})^{2.8}}{1+2.8(f/f_{p})^{3.8}}\,,
		\end{split}
	\end{equation}
	where $v_{w}$ is the bubble wall velocity and we simply choose $v_{w}=1$, $\kappa$ is the fraction of energy density transfered from vacuum into bubble walls, the subscript $*$ denotes the parameters evaluated at the PT time, i.e., $F(t_{*})=0.7$~\cite{Guth:1982cv,Ellis:2018mja}. The peak frequency $f_{p}$ reads
	\begin{equation}
		\begin{split}
			f_{p}=\frac{0.62}{1.8-0.1 v_{w}+v_{w}^{2}}\left(\beta/H_{*}\right)&\frac{T_{*}}{100\, \mathrm{GeV}}\times\\
			&1.65 \times 10^{-5}\,\mathrm{Hz}\,.
		\end{split}
	\end{equation}
	where $T_{*}$ is the PT temperature.
	
	To obtain the evolution of the energy density, we solve Eqs.~(\ref{eq:Gamma},\ref{eq:Ftn},\ref{eq:fried},\ref{eq:rho}) together self-consistently assuming $t_{i}=0$. In case 1, we choose $\beta/H_{*}=14.8$, $\alpha_{*}=6$ and $\kappa\rightarrow 1$. In case 2, we choose $\beta/H_{*}=3.7$, $\alpha_{*}=0.5$, $\kappa=1/3$. Fig.~\ref{fig:Frho} shows the evolution of the false vacuum fraction $F(t)$~(upper) and each component of the energy density~(lower) in case 1~(left) and case 2~(right), we can find that as $F(t)$ decreases, $\rho_{\mathrm{r}}(t)+\rho_{\mathrm{w}}(t)$ increases correspondingly, which means the vacuum energy tranfers into bubble walls and background plasma. Comparing the blue and orange lines in each panel of Fig.~\ref{fig:Frho}, the decrease of $F(t)$ is postponed for $t_{n}$, since $\rho_{r}$ scales as $a^{-4}$, the postpone of vacuum decay finally results in an increase of radiation energy density. At the end of simulation, $\rho_{\mathrm{inside}}/\rho_{\mathrm{outside}}$ exceeds the threshold $1+\delta_{c}=1.45$ and PBHs form.
	
	The PBH abundance depends on $\alpha_{*}$ and $\beta/H_{*}$.
	For smaller $\alpha_{*}$, Eq.~\eqref{eq:rho} implies the quantum decay in the overdense region should be postpone to a larger $t_{n}$ to reach the threshold $\delta_{c}$, then the probability $P(t_{n})$ decreases. For larger $\beta/H_{*}$, Eq.~\eqref{eq:prob} implies $P(t_{n})$ decreases with $t_{n}$ unchanged.
	Thus, the PBH abundance increases with $\alpha_{*}$ and decreases with $\beta/H_{*}$, while the PBH mass only depends on $T_{*}$ according to Eq.~\eqref{eq:PBHm}. 
	
	The PT temperature in case 1 and case 2 are chosen to be $T_{*}=1.58\times 10^{6}\,\mathrm{GeV}$ and $T_{*}=28.6\,\mathrm{MeV}$, which are inspired from Peccei-Quinn axion models~\cite{Peccei:1977ur} and hidden sector models~\cite{Chacko:2004ky}, respectively. In case 1 we choose $\gamma=1$, PBHs with mass $10^{20}$ g constitute all dark matter and the predicted GW energy spectrum peaks at $f_{p}=0.65\,\mathrm{Hz}$ with the peak value $\Omega_{\mathrm{GW},p}=4.3\times 10^{-9}$, which are expected to be detected by DECIGO, BBO, ET and CE, satisfying the current constraint by LIGO-Virgo collabration~\cite{Romero:2021kby}. In case 2 the predicted GW energy spectrum peaks at $f_{p}=4\times 10^{-9}\,\mathrm{Hz}$ with the peak value $\Omega_{\mathrm{GW},p}=10^{-9}$, which successfully explains the common spectrum process detected by NANOGrav collabration~\cite{Arzoumanian:2020vkk,Bian:2021lmz}.
	PBHs formed in case 2 with the parameter $\gamma=1/5$, whose abundance is predicted to be $\rho_{\mathrm{PBH}}/\rho_{\mathrm{DM}}=10^{-3}$, can explain the coalescence events observed by LIGO-Virgo collabration~\cite{Deng:2021ezy}.
	Fig.~\ref{fig:pbhgw} shows the predicted GW energy spectrum~(left) and PBH mass funcion~(right) in each case. Note that the nearly monochromatic mass function is an approximation here. In principle, one needs to consider the detailed collapsing process and the evolution of PBHs to obtain the small width of the mass function.

	\emph{Conclusion and discussion}.
	In this letter, we consider a novel mechanism of PBH formation during PTs as a model-independent universal consequence. Based on the numerical result of the collapse threshold of the whole Hubble horizon, we obtain an exact result of the predicted PBH abundance, and the mass function is almost monochromatic. We investigate two typical cases as examples and find that 1) PBHs from a PT could constitute all dark matter, and the predicted $\Omega_{\mathrm{GW}}$ peaks at about $0.65\,\mathrm{Hz}$, which is expected to be observed by DECIGO, BBO, CE, and ET. 2)GWs and PBHs produced from a PT could both explain the common-spectrum process observed by NANOGrav and the coalescence events observed by the LIGO-Virgo collaboration. 
	
	Based on the mechanism in this work, we can further realize a more accurate GW-PBH mutual constraint in the PT case. Roughly speaking, for the PT temperature $T_{*}<5\times 10^{8}$ GeV~(so that the PBHs are not evaporated), $\Omega_{\mathrm{GW}}<10^{-7}$ is ruled out by the overproduction of PBHs. Our mechanism can give constraints on PTs parameters at a higher energy scale comparing with that from the GW observation of LIGO-Virgo collaboration~\cite{Romero:2021kby}. The upcoming GW detectors can also give strict constraints on PBHs from PTs. 
	
	For the models where $v_{w}\lesssim 1$, the results remains roughly the same. In the other models where the bubble wall interacts strongly with background plasma, $v_{w}$ is then much smaller than the speed of light. We qualitatively conclude here that both the GW strength and the PBH abundance are largely reduced in this case, and leave the quantitative analysis for future work.
	Since the GW energy spectrum from PTs depends on other parameters such as $v_{w}$, the sound speed, and the interaction between bubble walls and background plasma, from GWs produced by bubble collisions and sound waves one cannot determine all of the PT parameters\footnote{From the analytical point of view, the envelope approximation~\cite{Kosowsky:1992vn,Weir:2016tov}, bulk flow model~\cite{Konstandin:2017sat,Jinno:2017fby} and sound shell method~\cite{Hindmarsh:2016lnk,Hindmarsh:2019phv} give the GW energy spectrum from bubble collisions and sound waves.
		Refs.~\cite{Hindmarsh:2013xza,Hindmarsh:2015qta,Cutting:2018tjt,Cutting:2020nla,Di:2020nny} conduct precise numerical simulations for GWs from PTs. The characteristic energy spectrum of GWs helps to distinguish PT models and determine model parameters~\cite{Caprini:2015zlo,Cai:2017tmh}.  }. Thus, the observations of both PBHs and GWs can eliminate degeneracy and give more strict constraints on model parameters.

	The PBH formation mechanism proposed in this letter does not suffer from the fine-tuning problem encountered in the case of PBH production from curvature perturbations. PBHs can be overproduced during strong and slow PTs which happened at a temperature $T_{*}>10^{9}$ GeV. These PBHs then dominate the Universe before their evaporation time, and the Planck relics of Hawking radiation could also account for dark matter. For the low-scale PTs inspired by the axion models, GWs may be detected through the B-mode polarization in the cosmic microwave background, and supermassive black holes with $10^6$ solar mass can also be produced for an $\mathcal{O}(100)$keV-scale dark PT.
	
	In this work, we consider the collapse of the overdense regions in the whole Hubble volume. However, small false vacuum bubbles are more common in the same situation. These bubbles tend to shrink into a very small region and the energy is then highly concentrated. The interactions become so violent that nonlinear evolution may arise in such a small region.  One needs numerical relativity simulations to obtain the final fate of the false vacuum bubbles. Moreover, smaller PBHs may arise if one takes into account the PBHs from bubble collisions surounding the overdense regions. Thus, the result of PBH mass function in this work is a conservative estimation and smaller PBHs may be more abundant. 
	
	
	We plan to consider the following aspects to obtain more accurate results. 1) The lattice simulations of PTs in the Friedmann-Robertson-Walker Universe can give more accurate results of $\Omega_{\mathrm{GW}}$. 2) In more realistic models, such as the results presented in numerical simulations~\cite{Hindmarsh:2015qta,Cutting:2018tjt,Cutting:2020nla}, the equation of state parameter is not exactly $1/3$ at the end of PT. This may be caused by the equation of state parameter of bubble walls for $v_{w}<c$, or the non-zero effective mass of the scalar field near the true vacuum. The threshold $\delta_{c}$ is smaller in this case as stated in Ref.~\cite{Musco:2012au}, which results in larger PBH abundance. 
	For this work's originality and physical meaning, these problems deserve more detailed research which we leave for future work.

	\emph{Acknowledgments}
	We thank Lang Liu and Da Huang for fruitful discussions.
	This work is supported in part by the National Key Research and Development Program of China Grants No. 2020YFC2201501 and No. 2021YFC2203004, in part by the National Natural Science Foundation of China Grants
	No. 11690021, No. 11690022, No. 11851302, No. 11947302, No. 11991052, No. 11821505, No. 12105060 and No. 12105344, in part by the Science Research Grants from the China Manned Space Project with NO. CMS-CSST-2021-B01,
	in part by the Strategic Priority Research Program of the Chinese Academy of Sciences Grant No. XDB23030100, in part by the Key Research Program of the CAS Grant No. XDPB15 and by Key Research Program of Frontier Sciences, CAS. Ligong Bian is supported by the National Natural Science Foundation of China under the grants Nos. 12075041, 12047564, and the Fundamental Research Funds for the Central Universities of China (No. 2021CDJQY-011 and No. 2020CDJQY-Z003),  and Chongqing Natural Science Foundation (Grants No.cstc2020jcyj-msxmX0814).
	\bibliography{PBHPT}

\begin{thebibliography}{78}%
\makeatletter
\providecommand \@ifxundefined [1]{%
 \@ifx{#1\undefined}
}%
\providecommand \@ifnum [1]{%
 \ifnum #1\expandafter \@firstoftwo
 \else \expandafter \@secondoftwo
 \fi
}%
\providecommand \@ifx [1]{%
 \ifx #1\expandafter \@firstoftwo
 \else \expandafter \@secondoftwo
 \fi
}%
\providecommand \natexlab [1]{#1}%
\providecommand \enquote  [1]{``#1''}%
\providecommand \bibnamefont  [1]{#1}%
\providecommand \bibfnamefont [1]{#1}%
\providecommand \citenamefont [1]{#1}%
\providecommand \href@noop [0]{\@secondoftwo}%
\providecommand \href [0]{\begingroup \@sanitize@url \@href}%
\providecommand \@href[1]{\@@startlink{#1}\@@href}%
\providecommand \@@href[1]{\endgroup#1\@@endlink}%
\providecommand \@sanitize@url [0]{\catcode `\\12\catcode `\$12\catcode
  `\&12\catcode `\#12\catcode `\^12\catcode `\_12\catcode `\%12\relax}%
\providecommand \@@startlink[1]{}%
\providecommand \@@endlink[0]{}%
\providecommand \url  [0]{\begingroup\@sanitize@url \@url }%
\providecommand \@url [1]{\endgroup\@href {#1}{\urlprefix }}%
\providecommand \urlprefix  [0]{URL }%
\providecommand \Eprint [0]{\href }%
\providecommand \doibase [0]{http://dx.doi.org/}%
\providecommand \selectlanguage [0]{\@gobble}%
\providecommand \bibinfo  [0]{\@secondoftwo}%
\providecommand \bibfield  [0]{\@secondoftwo}%
\providecommand \translation [1]{[#1]}%
\providecommand \BibitemOpen [0]{}%
\providecommand \bibitemStop [0]{}%
\providecommand \bibitemNoStop [0]{.\EOS\space}%
\providecommand \EOS [0]{\spacefactor3000\relax}%
\providecommand \BibitemShut  [1]{\csname bibitem#1\endcsname}%
\let\auto@bib@innerbib\@empty
\bibitem [{\citenamefont {Cai}\ \emph {et~al.}(2017{\natexlab{a}})\citenamefont
  {Cai}, \citenamefont {Cao}, \citenamefont {Guo}, \citenamefont {Wang},\ and\
  \citenamefont {Yang}}]{Cai:2017cbj}%
  \BibitemOpen
  \bibfield  {author} {\bibinfo {author} {\bibfnamefont {R.-G.}\ \bibnamefont
  {Cai}}, \bibinfo {author} {\bibfnamefont {Z.}~\bibnamefont {Cao}}, \bibinfo
  {author} {\bibfnamefont {Z.-K.}\ \bibnamefont {Guo}}, \bibinfo {author}
  {\bibfnamefont {S.-J.}\ \bibnamefont {Wang}}, \ and\ \bibinfo {author}
  {\bibfnamefont {T.}~\bibnamefont {Yang}},\ }\href {\doibase
  10.1093/nsr/nwx029} {\bibfield  {journal} {\bibinfo  {journal} {Natl. Sci.
  Rev.}\ }\textbf {\bibinfo {volume} {4}},\ \bibinfo {pages} {687} (\bibinfo
  {year} {2017}{\natexlab{a}})},\ \Eprint {http://arxiv.org/abs/1703.00187}
  {arXiv:1703.00187 [gr-qc]} \BibitemShut {NoStop}%
\bibitem [{\citenamefont {Bian}\ \emph
  {et~al.}(2021{\natexlab{a}})\citenamefont {Bian} \emph
  {et~al.}}]{Bian:2021ini}%
  \BibitemOpen
  \bibfield  {author} {\bibinfo {author} {\bibfnamefont {L.}~\bibnamefont
  {Bian}} \emph {et~al.},\ }\href@noop {} {\  (\bibinfo {year}
  {2021}{\natexlab{a}})},\ \Eprint {http://arxiv.org/abs/2106.10235}
  {arXiv:2106.10235 [gr-qc]} \BibitemShut {NoStop}%
\bibitem [{\citenamefont {Caprini}\ \emph {et~al.}(2020)\citenamefont {Caprini}
  \emph {et~al.}}]{Caprini:2019egz}%
  \BibitemOpen
  \bibfield  {author} {\bibinfo {author} {\bibfnamefont {C.}~\bibnamefont
  {Caprini}} \emph {et~al.},\ }\href {\doibase 10.1088/1475-7516/2020/03/024}
  {\bibfield  {journal} {\bibinfo  {journal} {JCAP}\ }\textbf {\bibinfo
  {volume} {03}},\ \bibinfo {pages} {024} (\bibinfo {year} {2020})},\ \Eprint
  {http://arxiv.org/abs/1910.13125} {arXiv:1910.13125 [astro-ph.CO]}
  \BibitemShut {NoStop}%
\bibitem [{\citenamefont {Coleman}(1977)}]{Coleman:1977py}%
  \BibitemOpen
  \bibfield  {author} {\bibinfo {author} {\bibfnamefont {S.~R.}\ \bibnamefont
  {Coleman}},\ }\href {\doibase 10.1103/PhysRevD.15.2929,
  10.1103/PhysRevD.16.1248} {\bibfield  {journal} {\bibinfo  {journal} {Phys.
  Rev.}\ }\textbf {\bibinfo {volume} {D15}},\ \bibinfo {pages} {2929} (\bibinfo
  {year} {1977})},\ \bibinfo {note} {[Erratum: Phys.
  Rev.D16,1248(1977)]}\BibitemShut {NoStop}%
\bibitem [{\citenamefont {Callan}\ and\ \citenamefont
  {Coleman}(1977)}]{Callan:1977pt}%
  \BibitemOpen
  \bibfield  {author} {\bibinfo {author} {\bibfnamefont {C.~G.}\ \bibnamefont
  {Callan}, \bibfnamefont {Jr.}}\ and\ \bibinfo {author} {\bibfnamefont
  {S.~R.}\ \bibnamefont {Coleman}},\ }\href {\doibase 10.1103/PhysRevD.16.1762}
  {\bibfield  {journal} {\bibinfo  {journal} {Phys. Rev.}\ }\textbf {\bibinfo
  {volume} {D16}},\ \bibinfo {pages} {1762} (\bibinfo {year}
  {1977})}\BibitemShut {NoStop}%
\bibitem [{\citenamefont {Espinosa}\ \emph {et~al.}(2010)\citenamefont
  {Espinosa}, \citenamefont {Konstandin}, \citenamefont {No},\ and\
  \citenamefont {Servant}}]{Espinosa:2010hh}%
  \BibitemOpen
  \bibfield  {author} {\bibinfo {author} {\bibfnamefont {J.~R.}\ \bibnamefont
  {Espinosa}}, \bibinfo {author} {\bibfnamefont {T.}~\bibnamefont
  {Konstandin}}, \bibinfo {author} {\bibfnamefont {J.~M.}\ \bibnamefont {No}},
  \ and\ \bibinfo {author} {\bibfnamefont {G.}~\bibnamefont {Servant}},\ }\href
  {\doibase 10.1088/1475-7516/2010/06/028} {\bibfield  {journal} {\bibinfo
  {journal} {JCAP}\ }\textbf {\bibinfo {volume} {06}},\ \bibinfo {pages} {028}
  (\bibinfo {year} {2010})},\ \Eprint {http://arxiv.org/abs/1004.4187}
  {arXiv:1004.4187 [hep-ph]} \BibitemShut {NoStop}%
\bibitem [{\citenamefont {Ellis}\ \emph
  {et~al.}(2019{\natexlab{a}})\citenamefont {Ellis}, \citenamefont {Lewicki},
  \citenamefont {No},\ and\ \citenamefont {Vaskonen}}]{Ellis:2019oqb}%
  \BibitemOpen
  \bibfield  {author} {\bibinfo {author} {\bibfnamefont {J.}~\bibnamefont
  {Ellis}}, \bibinfo {author} {\bibfnamefont {M.}~\bibnamefont {Lewicki}},
  \bibinfo {author} {\bibfnamefont {J.~M.}\ \bibnamefont {No}}, \ and\ \bibinfo
  {author} {\bibfnamefont {V.}~\bibnamefont {Vaskonen}},\ }\href {\doibase
  10.1088/1475-7516/2019/06/024} {\bibfield  {journal} {\bibinfo  {journal}
  {JCAP}\ }\textbf {\bibinfo {volume} {06}},\ \bibinfo {pages} {024} (\bibinfo
  {year} {2019}{\natexlab{a}})},\ \Eprint {http://arxiv.org/abs/1903.09642}
  {arXiv:1903.09642 [hep-ph]} \BibitemShut {NoStop}%
\bibitem [{\citenamefont {Cai}\ and\ \citenamefont {Wang}(2021)}]{Cai:2020djd}%
  \BibitemOpen
  \bibfield  {author} {\bibinfo {author} {\bibfnamefont {R.-G.}\ \bibnamefont
  {Cai}}\ and\ \bibinfo {author} {\bibfnamefont {S.-J.}\ \bibnamefont {Wang}},\
  }\href {\doibase 10.1088/1475-7516/2021/03/096} {\bibfield  {journal}
  {\bibinfo  {journal} {JCAP}\ }\textbf {\bibinfo {volume} {03}},\ \bibinfo
  {pages} {096} (\bibinfo {year} {2021})},\ \Eprint
  {http://arxiv.org/abs/2011.11451} {arXiv:2011.11451 [astro-ph.CO]}
  \BibitemShut {NoStop}%
\bibitem [{\citenamefont {Hawking}\ \emph {et~al.}(1982)\citenamefont
  {Hawking}, \citenamefont {Moss},\ and\ \citenamefont
  {Stewart}}]{Hawking:1982ga}%
  \BibitemOpen
  \bibfield  {author} {\bibinfo {author} {\bibfnamefont {S.~W.}\ \bibnamefont
  {Hawking}}, \bibinfo {author} {\bibfnamefont {I.~G.}\ \bibnamefont {Moss}}, \
  and\ \bibinfo {author} {\bibfnamefont {J.~M.}\ \bibnamefont {Stewart}},\
  }\href {\doibase 10.1103/PhysRevD.26.2681} {\bibfield  {journal} {\bibinfo
  {journal} {Phys. Rev. D}\ }\textbf {\bibinfo {volume} {26}},\ \bibinfo
  {pages} {2681} (\bibinfo {year} {1982})}\BibitemShut {NoStop}%
\bibitem [{\citenamefont {Crawford}\ and\ \citenamefont
  {Schramm}(1982)}]{Crawford:1982yz}%
  \BibitemOpen
  \bibfield  {author} {\bibinfo {author} {\bibfnamefont {M.}~\bibnamefont
  {Crawford}}\ and\ \bibinfo {author} {\bibfnamefont {D.~N.}\ \bibnamefont
  {Schramm}},\ }\href {\doibase 10.1038/298538a0} {\bibfield  {journal}
  {\bibinfo  {journal} {Nature}\ }\textbf {\bibinfo {volume} {298}},\ \bibinfo
  {pages} {538} (\bibinfo {year} {1982})}\BibitemShut {NoStop}%
\bibitem [{\citenamefont {Kodama}\ \emph {et~al.}(1982)\citenamefont {Kodama},
  \citenamefont {Sasaki},\ and\ \citenamefont {Sato}}]{Kodama:1982sf}%
  \BibitemOpen
  \bibfield  {author} {\bibinfo {author} {\bibfnamefont {H.}~\bibnamefont
  {Kodama}}, \bibinfo {author} {\bibfnamefont {M.}~\bibnamefont {Sasaki}}, \
  and\ \bibinfo {author} {\bibfnamefont {K.}~\bibnamefont {Sato}},\ }\href
  {\doibase 10.1143/PTP.68.1979} {\bibfield  {journal} {\bibinfo  {journal}
  {Prog. Theor. Phys.}\ }\textbf {\bibinfo {volume} {68}},\ \bibinfo {pages}
  {1979} (\bibinfo {year} {1982})}\BibitemShut {NoStop}%
\bibitem [{\citenamefont {Johnson}\ \emph {et~al.}(2012)\citenamefont
  {Johnson}, \citenamefont {Peiris},\ and\ \citenamefont
  {Lehner}}]{Johnson:2011wt}%
  \BibitemOpen
  \bibfield  {author} {\bibinfo {author} {\bibfnamefont {M.~C.}\ \bibnamefont
  {Johnson}}, \bibinfo {author} {\bibfnamefont {H.~V.}\ \bibnamefont {Peiris}},
  \ and\ \bibinfo {author} {\bibfnamefont {L.}~\bibnamefont {Lehner}},\ }\href
  {\doibase 10.1103/PhysRevD.85.083516} {\bibfield  {journal} {\bibinfo
  {journal} {Phys. Rev.}\ }\textbf {\bibinfo {volume} {D85}},\ \bibinfo {pages}
  {083516} (\bibinfo {year} {2012})},\ \Eprint {http://arxiv.org/abs/1112.4487}
  {arXiv:1112.4487 [hep-th]} \BibitemShut {NoStop}%
\bibitem [{\citenamefont {Khlopov}\ \emph {et~al.}(1998)\citenamefont
  {Khlopov}, \citenamefont {Konoplich}, \citenamefont {Rubin},\ and\
  \citenamefont {Sakharov}}]{Khlopov:1998nm}%
  \BibitemOpen
  \bibfield  {author} {\bibinfo {author} {\bibfnamefont {M.~{\relax Yu}.}\
  \bibnamefont {Khlopov}}, \bibinfo {author} {\bibfnamefont {R.~V.}\
  \bibnamefont {Konoplich}}, \bibinfo {author} {\bibfnamefont {S.~G.}\
  \bibnamefont {Rubin}}, \ and\ \bibinfo {author} {\bibfnamefont {A.~S.}\
  \bibnamefont {Sakharov}},\ }\href@noop {} {\  (\bibinfo {year} {1998})},\
  \Eprint {http://arxiv.org/abs/hep-ph/9807343} {arXiv:hep-ph/9807343 [hep-ph]}
  \BibitemShut {NoStop}%
\bibitem [{\citenamefont {Carr}\ \emph {et~al.}(2016)\citenamefont {Carr},
  \citenamefont {Kuhnel},\ and\ \citenamefont {Sandstad}}]{Carr:2016drx}%
  \BibitemOpen
  \bibfield  {author} {\bibinfo {author} {\bibfnamefont {B.}~\bibnamefont
  {Carr}}, \bibinfo {author} {\bibfnamefont {F.}~\bibnamefont {Kuhnel}}, \ and\
  \bibinfo {author} {\bibfnamefont {M.}~\bibnamefont {Sandstad}},\ }\href
  {\doibase 10.1103/PhysRevD.94.083504} {\bibfield  {journal} {\bibinfo
  {journal} {Phys. Rev.}\ }\textbf {\bibinfo {volume} {D94}},\ \bibinfo {pages}
  {083504} (\bibinfo {year} {2016})},\ \Eprint
  {http://arxiv.org/abs/1607.06077} {arXiv:1607.06077 [astro-ph.CO]}
  \BibitemShut {NoStop}%
\bibitem [{\citenamefont {Carr}\ and\ \citenamefont
  {Kuhnel}(2020)}]{Carr:2020xqk}%
  \BibitemOpen
  \bibfield  {author} {\bibinfo {author} {\bibfnamefont {B.}~\bibnamefont
  {Carr}}\ and\ \bibinfo {author} {\bibfnamefont {F.}~\bibnamefont {Kuhnel}},\
  }\href {\doibase 10.1146/annurev-nucl-050520-125911} {\bibfield  {journal}
  {\bibinfo  {journal} {Ann. Rev. Nucl. Part. Sci.}\ }\textbf {\bibinfo
  {volume} {70}},\ \bibinfo {pages} {355} (\bibinfo {year} {2020})},\ \Eprint
  {http://arxiv.org/abs/2006.02838} {arXiv:2006.02838 [astro-ph.CO]}
  \BibitemShut {NoStop}%
\bibitem [{\citenamefont {Sasaki}\ \emph {et~al.}(2016)\citenamefont {Sasaki},
  \citenamefont {Suyama}, \citenamefont {Tanaka},\ and\ \citenamefont
  {Yokoyama}}]{Sasaki:2016jop}%
  \BibitemOpen
  \bibfield  {author} {\bibinfo {author} {\bibfnamefont {M.}~\bibnamefont
  {Sasaki}}, \bibinfo {author} {\bibfnamefont {T.}~\bibnamefont {Suyama}},
  \bibinfo {author} {\bibfnamefont {T.}~\bibnamefont {Tanaka}}, \ and\ \bibinfo
  {author} {\bibfnamefont {S.}~\bibnamefont {Yokoyama}},\ }\href {\doibase
  10.1103/PhysRevLett.121.059901, 10.1103/PhysRevLett.117.061101} {\bibfield
  {journal} {\bibinfo  {journal} {Phys. Rev. Lett.}\ }\textbf {\bibinfo
  {volume} {117}},\ \bibinfo {pages} {061101} (\bibinfo {year} {2016})},\
  \bibinfo {note} {[erratum: Phys. Rev. Lett.121,no.5,059901(2018)]},\ \Eprint
  {http://arxiv.org/abs/1603.08338} {arXiv:1603.08338 [astro-ph.CO]}
  \BibitemShut {NoStop}%
\bibitem [{\citenamefont {Bird}\ \emph {et~al.}(2016)\citenamefont {Bird},
  \citenamefont {Cholis}, \citenamefont {Muñoz}, \citenamefont {Ali-Haïmoud},
  \citenamefont {Kamionkowski}, \citenamefont {Kovetz}, \citenamefont
  {Raccanelli},\ and\ \citenamefont {Riess}}]{Bird:2016dcv}%
  \BibitemOpen
  \bibfield  {author} {\bibinfo {author} {\bibfnamefont {S.}~\bibnamefont
  {Bird}}, \bibinfo {author} {\bibfnamefont {I.}~\bibnamefont {Cholis}},
  \bibinfo {author} {\bibfnamefont {J.~B.}\ \bibnamefont {Muñoz}}, \bibinfo
  {author} {\bibfnamefont {Y.}~\bibnamefont {Ali-Haïmoud}}, \bibinfo {author}
  {\bibfnamefont {M.}~\bibnamefont {Kamionkowski}}, \bibinfo {author}
  {\bibfnamefont {E.~D.}\ \bibnamefont {Kovetz}}, \bibinfo {author}
  {\bibfnamefont {A.}~\bibnamefont {Raccanelli}}, \ and\ \bibinfo {author}
  {\bibfnamefont {A.~G.}\ \bibnamefont {Riess}},\ }\href {\doibase
  10.1103/PhysRevLett.116.201301} {\bibfield  {journal} {\bibinfo  {journal}
  {Phys. Rev. Lett.}\ }\textbf {\bibinfo {volume} {116}},\ \bibinfo {pages}
  {201301} (\bibinfo {year} {2016})},\ \Eprint
  {http://arxiv.org/abs/1603.00464} {arXiv:1603.00464 [astro-ph.CO]}
  \BibitemShut {NoStop}%
\bibitem [{\citenamefont {Niemeyer}\ and\ \citenamefont
  {Jedamzik}(1998)}]{Niemeyer:1997mt}%
  \BibitemOpen
  \bibfield  {author} {\bibinfo {author} {\bibfnamefont {J.~C.}\ \bibnamefont
  {Niemeyer}}\ and\ \bibinfo {author} {\bibfnamefont {K.}~\bibnamefont
  {Jedamzik}},\ }\href {\doibase 10.1103/PhysRevLett.80.5481} {\bibfield
  {journal} {\bibinfo  {journal} {Phys. Rev. Lett.}\ }\textbf {\bibinfo
  {volume} {80}},\ \bibinfo {pages} {5481} (\bibinfo {year} {1998})},\ \Eprint
  {http://arxiv.org/abs/astro-ph/9709072} {arXiv:astro-ph/9709072} \BibitemShut
  {NoStop}%
\bibitem [{\citenamefont {K\"uhnel}\ \emph {et~al.}(2016)\citenamefont
  {K\"uhnel}, \citenamefont {Rampf},\ and\ \citenamefont
  {Sandstad}}]{Kuhnel:2015vtw}%
  \BibitemOpen
  \bibfield  {author} {\bibinfo {author} {\bibfnamefont {F.}~\bibnamefont
  {K\"uhnel}}, \bibinfo {author} {\bibfnamefont {C.}~\bibnamefont {Rampf}}, \
  and\ \bibinfo {author} {\bibfnamefont {M.}~\bibnamefont {Sandstad}},\ }\href
  {\doibase 10.1140/epjc/s10052-016-3945-8} {\bibfield  {journal} {\bibinfo
  {journal} {Eur. Phys. J. C}\ }\textbf {\bibinfo {volume} {76}},\ \bibinfo
  {pages} {93} (\bibinfo {year} {2016})},\ \Eprint
  {http://arxiv.org/abs/1512.00488} {arXiv:1512.00488 [astro-ph.CO]}
  \BibitemShut {NoStop}%
\bibitem [{\citenamefont {Gross}\ \emph {et~al.}(2021)\citenamefont {Gross},
  \citenamefont {Landini}, \citenamefont {Strumia},\ and\ \citenamefont
  {Teresi}}]{Gross:2021qgx}%
  \BibitemOpen
  \bibfield  {author} {\bibinfo {author} {\bibfnamefont {C.}~\bibnamefont
  {Gross}}, \bibinfo {author} {\bibfnamefont {G.}~\bibnamefont {Landini}},
  \bibinfo {author} {\bibfnamefont {A.}~\bibnamefont {Strumia}}, \ and\
  \bibinfo {author} {\bibfnamefont {D.}~\bibnamefont {Teresi}},\ }\href@noop {}
  {\  (\bibinfo {year} {2021})},\ \Eprint {http://arxiv.org/abs/2105.02840}
  {arXiv:2105.02840 [hep-ph]} \BibitemShut {NoStop}%
\bibitem [{\citenamefont {Baker}\ \emph {et~al.}(2021)\citenamefont {Baker},
  \citenamefont {Breitbach}, \citenamefont {Kopp},\ and\ \citenamefont
  {Mittnacht}}]{Baker:2021nyl}%
  \BibitemOpen
  \bibfield  {author} {\bibinfo {author} {\bibfnamefont {M.~J.}\ \bibnamefont
  {Baker}}, \bibinfo {author} {\bibfnamefont {M.}~\bibnamefont {Breitbach}},
  \bibinfo {author} {\bibfnamefont {J.}~\bibnamefont {Kopp}}, \ and\ \bibinfo
  {author} {\bibfnamefont {L.}~\bibnamefont {Mittnacht}},\ }\href@noop {} {\
  (\bibinfo {year} {2021})},\ \Eprint {http://arxiv.org/abs/2105.07481}
  {arXiv:2105.07481 [astro-ph.CO]} \BibitemShut {NoStop}%
\bibitem [{\citenamefont {Kawana}\ and\ \citenamefont
  {Xie}(2021)}]{Kawana:2021tde}%
  \BibitemOpen
  \bibfield  {author} {\bibinfo {author} {\bibfnamefont {K.}~\bibnamefont
  {Kawana}}\ and\ \bibinfo {author} {\bibfnamefont {K.-P.}\ \bibnamefont
  {Xie}},\ }\href@noop {} {\  (\bibinfo {year} {2021})},\ \Eprint
  {http://arxiv.org/abs/2106.00111} {arXiv:2106.00111 [astro-ph.CO]}
  \BibitemShut {NoStop}%
\bibitem [{\citenamefont {Musco}\ \emph {et~al.}(2005)\citenamefont {Musco},
  \citenamefont {Miller},\ and\ \citenamefont {Rezzolla}}]{Musco:2004ak}%
  \BibitemOpen
  \bibfield  {author} {\bibinfo {author} {\bibfnamefont {I.}~\bibnamefont
  {Musco}}, \bibinfo {author} {\bibfnamefont {J.~C.}\ \bibnamefont {Miller}}, \
  and\ \bibinfo {author} {\bibfnamefont {L.}~\bibnamefont {Rezzolla}},\ }\href
  {\doibase 10.1088/0264-9381/22/7/013} {\bibfield  {journal} {\bibinfo
  {journal} {Class. Quant. Grav.}\ }\textbf {\bibinfo {volume} {22}},\ \bibinfo
  {pages} {1405} (\bibinfo {year} {2005})},\ \Eprint
  {http://arxiv.org/abs/gr-qc/0412063} {arXiv:gr-qc/0412063 [gr-qc]}
  \BibitemShut {NoStop}%
\bibitem [{\citenamefont {Harada}\ \emph {et~al.}(2013)\citenamefont {Harada},
  \citenamefont {Yoo},\ and\ \citenamefont {Kohri}}]{Harada:2013epa}%
  \BibitemOpen
  \bibfield  {author} {\bibinfo {author} {\bibfnamefont {T.}~\bibnamefont
  {Harada}}, \bibinfo {author} {\bibfnamefont {C.-M.}\ \bibnamefont {Yoo}}, \
  and\ \bibinfo {author} {\bibfnamefont {K.}~\bibnamefont {Kohri}},\ }\href
  {\doibase 10.1103/PhysRevD.88.084051, 10.1103/PhysRevD.89.029903} {\bibfield
  {journal} {\bibinfo  {journal} {Phys. Rev.}\ }\textbf {\bibinfo {volume}
  {D88}},\ \bibinfo {pages} {084051} (\bibinfo {year} {2013})},\ \bibinfo
  {note} {[Erratum: Phys. Rev.D89,no.2,029903(2014)]},\ \Eprint
  {http://arxiv.org/abs/1309.4201} {arXiv:1309.4201 [astro-ph.CO]} \BibitemShut
  {NoStop}%
\bibitem [{\citenamefont {Enqvist}\ \emph {et~al.}(1992)\citenamefont
  {Enqvist}, \citenamefont {Ignatius}, \citenamefont {Kajantie},\ and\
  \citenamefont {Rummukainen}}]{Enqvist:1991xw}%
  \BibitemOpen
  \bibfield  {author} {\bibinfo {author} {\bibfnamefont {K.}~\bibnamefont
  {Enqvist}}, \bibinfo {author} {\bibfnamefont {J.}~\bibnamefont {Ignatius}},
  \bibinfo {author} {\bibfnamefont {K.}~\bibnamefont {Kajantie}}, \ and\
  \bibinfo {author} {\bibfnamefont {K.}~\bibnamefont {Rummukainen}},\ }\href
  {\doibase 10.1103/PhysRevD.45.3415} {\bibfield  {journal} {\bibinfo
  {journal} {Phys. Rev. D}\ }\textbf {\bibinfo {volume} {45}},\ \bibinfo
  {pages} {3415} (\bibinfo {year} {1992})}\BibitemShut {NoStop}%
\bibitem [{\citenamefont {Turner}\ \emph {et~al.}(1992)\citenamefont {Turner},
  \citenamefont {Weinberg},\ and\ \citenamefont {Widrow}}]{Turner:1992tz}%
  \BibitemOpen
  \bibfield  {author} {\bibinfo {author} {\bibfnamefont {M.~S.}\ \bibnamefont
  {Turner}}, \bibinfo {author} {\bibfnamefont {E.~J.}\ \bibnamefont
  {Weinberg}}, \ and\ \bibinfo {author} {\bibfnamefont {L.~M.}\ \bibnamefont
  {Widrow}},\ }\href {\doibase 10.1103/PhysRevD.46.2384} {\bibfield  {journal}
  {\bibinfo  {journal} {Phys. Rev.}\ }\textbf {\bibinfo {volume} {D46}},\
  \bibinfo {pages} {2384} (\bibinfo {year} {1992})}\BibitemShut {NoStop}%
\bibitem [{\citenamefont {Musco}\ \emph {et~al.}(2021)\citenamefont {Musco},
  \citenamefont {De~Luca}, \citenamefont {Franciolini},\ and\ \citenamefont
  {Riotto}}]{Musco:2020jjb}%
  \BibitemOpen
  \bibfield  {author} {\bibinfo {author} {\bibfnamefont {I.}~\bibnamefont
  {Musco}}, \bibinfo {author} {\bibfnamefont {V.}~\bibnamefont {De~Luca}},
  \bibinfo {author} {\bibfnamefont {G.}~\bibnamefont {Franciolini}}, \ and\
  \bibinfo {author} {\bibfnamefont {A.}~\bibnamefont {Riotto}},\ }\href
  {\doibase 10.1103/PhysRevD.103.063538} {\bibfield  {journal} {\bibinfo
  {journal} {Phys. Rev.}\ }\textbf {\bibinfo {volume} {D103}},\ \bibinfo
  {pages} {063538} (\bibinfo {year} {2021})},\ \Eprint
  {http://arxiv.org/abs/2011.03014} {arXiv:2011.03014 [astro-ph.CO]}
  \BibitemShut {NoStop}%
\bibitem [{\citenamefont {Schmitz}(2021)}]{Schmitz:2020syl}%
  \BibitemOpen
  \bibfield  {author} {\bibinfo {author} {\bibfnamefont {K.}~\bibnamefont
  {Schmitz}},\ }\href {\doibase 10.1007/JHEP01(2021)097} {\bibfield  {journal}
  {\bibinfo  {journal} {JHEP}\ }\textbf {\bibinfo {volume} {01}},\ \bibinfo
  {pages} {097} (\bibinfo {year} {2021})},\ \Eprint
  {http://arxiv.org/abs/2002.04615} {arXiv:2002.04615 [hep-ph]} \BibitemShut
  {NoStop}%
\bibitem [{\citenamefont {Lentati}\ \emph {et~al.}(2015)\citenamefont {Lentati}
  \emph {et~al.}}]{Lentati:2015qwp}%
  \BibitemOpen
  \bibfield  {author} {\bibinfo {author} {\bibfnamefont {L.}~\bibnamefont
  {Lentati}} \emph {et~al.},\ }\href {\doibase 10.1093/mnras/stv1538}
  {\bibfield  {journal} {\bibinfo  {journal} {Mon. Not. Roy. Astron. Soc.}\
  }\textbf {\bibinfo {volume} {453}},\ \bibinfo {pages} {2576} (\bibinfo {year}
  {2015})},\ \Eprint {http://arxiv.org/abs/1504.03692} {arXiv:1504.03692
  [astro-ph.CO]} \BibitemShut {NoStop}%
\bibitem [{\citenamefont {Shannon}\ \emph {et~al.}(2015)\citenamefont {Shannon}
  \emph {et~al.}}]{Shannon:2015ect}%
  \BibitemOpen
  \bibfield  {author} {\bibinfo {author} {\bibfnamefont {R.~M.}\ \bibnamefont
  {Shannon}} \emph {et~al.},\ }\href {\doibase 10.1126/science.aab1910}
  {\bibfield  {journal} {\bibinfo  {journal} {Science}\ }\textbf {\bibinfo
  {volume} {349}},\ \bibinfo {pages} {1522} (\bibinfo {year} {2015})},\ \Eprint
  {http://arxiv.org/abs/1509.07320} {arXiv:1509.07320 [astro-ph.CO]}
  \BibitemShut {NoStop}%
\bibitem [{\citenamefont {Arzoumanian}\ \emph {et~al.}(2018)\citenamefont
  {Arzoumanian} \emph {et~al.}}]{Arzoumanian:2018saf}%
  \BibitemOpen
  \bibfield  {author} {\bibinfo {author} {\bibfnamefont {Z.}~\bibnamefont
  {Arzoumanian}} \emph {et~al.} (\bibinfo {collaboration} {NANOGRAV}),\ }\href
  {\doibase 10.3847/1538-4357/aabd3b} {\bibfield  {journal} {\bibinfo
  {journal} {Astrophys. J.}\ }\textbf {\bibinfo {volume} {859}},\ \bibinfo
  {pages} {47} (\bibinfo {year} {2018})},\ \Eprint
  {http://arxiv.org/abs/1801.02617} {arXiv:1801.02617 [astro-ph.HE]}
  \BibitemShut {NoStop}%
\bibitem [{\citenamefont {Hobbs}\ \emph {et~al.}(2010)\citenamefont {Hobbs}
  \emph {et~al.}}]{Hobbs:2009yy}%
  \BibitemOpen
  \bibfield  {author} {\bibinfo {author} {\bibfnamefont {G.}~\bibnamefont
  {Hobbs}} \emph {et~al.},\ }\bibfield  {booktitle} {\emph {\bibinfo
  {booktitle} {{Gravitational waves. Proceedings, 8th Edoardo Amaldi
  Conference, Amaldi 8, New York, USA, June 22-26, 2009}}},\ }\href {\doibase
  10.1088/0264-9381/27/8/084013} {\bibfield  {journal} {\bibinfo  {journal}
  {Class. Quant. Grav.}\ }\textbf {\bibinfo {volume} {27}},\ \bibinfo {pages}
  {084013} (\bibinfo {year} {2010})},\ \Eprint {http://arxiv.org/abs/0911.5206}
  {arXiv:0911.5206 [astro-ph.SR]} \BibitemShut {NoStop}%
\bibitem [{\citenamefont {Carilli}\ and\ \citenamefont
  {Rawlings}(2004)}]{Carilli:2004nx}%
  \BibitemOpen
  \bibfield  {author} {\bibinfo {author} {\bibfnamefont {C.~L.}\ \bibnamefont
  {Carilli}}\ and\ \bibinfo {author} {\bibfnamefont {S.}~\bibnamefont
  {Rawlings}},\ }\bibfield  {booktitle} {\emph {\bibinfo {booktitle}
  {{International SKA Conference 2003 Geraldton, Australia, July 27-August 2,
  2003}}},\ }\href {\doibase 10.1016/j.newar.2004.09.001} {\bibfield  {journal}
  {\bibinfo  {journal} {New Astron. Rev.}\ }\textbf {\bibinfo {volume} {48}},\
  \bibinfo {pages} {979} (\bibinfo {year} {2004})},\ \Eprint
  {http://arxiv.org/abs/astro-ph/0409274} {arXiv:astro-ph/0409274 [astro-ph]}
  \BibitemShut {NoStop}%
\bibitem [{\citenamefont {Amaro-Seoane}\ \emph {et~al.}(2017)\citenamefont
  {Amaro-Seoane} \emph {et~al.}}]{Audley:2017drz}%
  \BibitemOpen
  \bibfield  {author} {\bibinfo {author} {\bibfnamefont {P.}~\bibnamefont
  {Amaro-Seoane}} \emph {et~al.} (\bibinfo {collaboration} {LISA}),\
  }\href@noop {} {\  (\bibinfo {year} {2017})},\ \Eprint
  {http://arxiv.org/abs/1702.00786} {arXiv:1702.00786 [astro-ph.IM]}
  \BibitemShut {NoStop}%
\bibitem [{\citenamefont {Ruan}\ \emph {et~al.}(2018)\citenamefont {Ruan},
  \citenamefont {Guo}, \citenamefont {Cai},\ and\ \citenamefont
  {Zhang}}]{Guo:2018npi}%
  \BibitemOpen
  \bibfield  {author} {\bibinfo {author} {\bibfnamefont {W.-H.}\ \bibnamefont
  {Ruan}}, \bibinfo {author} {\bibfnamefont {Z.-K.}\ \bibnamefont {Guo}},
  \bibinfo {author} {\bibfnamefont {R.-G.}\ \bibnamefont {Cai}}, \ and\
  \bibinfo {author} {\bibfnamefont {Y.-Z.}\ \bibnamefont {Zhang}},\ }\href@noop
  {} {\  (\bibinfo {year} {2018})},\ \Eprint {http://arxiv.org/abs/1807.09495}
  {arXiv:1807.09495 [gr-qc]} \BibitemShut {NoStop}%
\bibitem [{\citenamefont {Kawamura}\ \emph {et~al.}(2011)\citenamefont
  {Kawamura} \emph {et~al.}}]{Kawamura:2011zz}%
  \BibitemOpen
  \bibfield  {author} {\bibinfo {author} {\bibfnamefont {S.}~\bibnamefont
  {Kawamura}} \emph {et~al.},\ }\bibfield  {booktitle} {\emph {\bibinfo
  {booktitle} {{Laser interferometer space antenna. Proceedings, 8th
  International LISA Symposium, Stanford, USA, June 28-July 2, 2010}}},\ }\href
  {\doibase 10.1088/0264-9381/28/9/094011} {\bibfield  {journal} {\bibinfo
  {journal} {Class. Quant. Grav.}\ }\textbf {\bibinfo {volume} {28}},\ \bibinfo
  {pages} {094011} (\bibinfo {year} {2011})}\BibitemShut {NoStop}%
\bibitem [{\citenamefont {Phinney}\ \emph {et~al.}(2004)\citenamefont
  {Phinney}, \citenamefont {Bender}, \citenamefont {Buchman}, \citenamefont
  {Byer}, \citenamefont {Cornish}, \citenamefont {Fritschel}, \citenamefont
  {Folkner}, \citenamefont {Merkowitz}, \citenamefont {Danzmann}, \citenamefont
  {DiFiore} \emph {et~al.}}]{phinney2004big}%
  \BibitemOpen
  \bibfield  {author} {\bibinfo {author} {\bibfnamefont {S.}~\bibnamefont
  {Phinney}}, \bibinfo {author} {\bibfnamefont {P.}~\bibnamefont {Bender}},
  \bibinfo {author} {\bibfnamefont {R.}~\bibnamefont {Buchman}}, \bibinfo
  {author} {\bibfnamefont {R.}~\bibnamefont {Byer}}, \bibinfo {author}
  {\bibfnamefont {N.}~\bibnamefont {Cornish}}, \bibinfo {author} {\bibfnamefont
  {P.}~\bibnamefont {Fritschel}}, \bibinfo {author} {\bibfnamefont
  {W.}~\bibnamefont {Folkner}}, \bibinfo {author} {\bibfnamefont
  {S.}~\bibnamefont {Merkowitz}}, \bibinfo {author} {\bibfnamefont
  {K.}~\bibnamefont {Danzmann}}, \bibinfo {author} {\bibfnamefont
  {L.}~\bibnamefont {DiFiore}},  \emph {et~al.},\ }\href@noop {} {\bibfield
  {journal} {\bibinfo  {journal} {NASA Mission Concept Study}\ } (\bibinfo
  {year} {2004})}\BibitemShut {NoStop}%
\bibitem [{\citenamefont {Aasi}\ \emph {et~al.}(2015)\citenamefont {Aasi} \emph
  {et~al.}}]{TheLIGOScientific:2014jea}%
  \BibitemOpen
  \bibfield  {author} {\bibinfo {author} {\bibfnamefont {J.}~\bibnamefont
  {Aasi}} \emph {et~al.} (\bibinfo {collaboration} {LIGO Scientific}),\ }\href
  {\doibase 10.1088/0264-9381/32/7/074001} {\bibfield  {journal} {\bibinfo
  {journal} {Class. Quant. Grav.}\ }\textbf {\bibinfo {volume} {32}},\ \bibinfo
  {pages} {074001} (\bibinfo {year} {2015})},\ \Eprint
  {http://arxiv.org/abs/1411.4547} {arXiv:1411.4547 [gr-qc]} \BibitemShut
  {NoStop}%
\bibitem [{\citenamefont {Somiya}(2012)}]{Somiya:2011np}%
  \BibitemOpen
  \bibfield  {author} {\bibinfo {author} {\bibfnamefont {K.}~\bibnamefont
  {Somiya}} (\bibinfo {collaboration} {KAGRA}),\ }\bibfield  {booktitle} {\emph
  {\bibinfo {booktitle} {{Gravitational waves. Numerical relativity - data
  analysis. Proceedings, 9th Edoardo Amaldi Conference, Amaldi 9, and meeting,
  NRDA 2011, Cardiff, UK, July 10-15, 2011}}},\ }\href {\doibase
  10.1088/0264-9381/29/12/124007} {\bibfield  {journal} {\bibinfo  {journal}
  {Class. Quant. Grav.}\ }\textbf {\bibinfo {volume} {29}},\ \bibinfo {pages}
  {124007} (\bibinfo {year} {2012})},\ \Eprint {http://arxiv.org/abs/1111.7185}
  {arXiv:1111.7185 [gr-qc]} \BibitemShut {NoStop}%
\bibitem [{\citenamefont {Reitze}\ \emph {et~al.}(2019)\citenamefont {Reitze}
  \emph {et~al.}}]{Reitze:2019iox}%
  \BibitemOpen
  \bibfield  {author} {\bibinfo {author} {\bibfnamefont {D.}~\bibnamefont
  {Reitze}} \emph {et~al.},\ }\href@noop {} {\bibfield  {journal} {\bibinfo
  {journal} {Bull. Am. Astron. Soc.}\ }\textbf {\bibinfo {volume} {51}},\
  \bibinfo {pages} {035} (\bibinfo {year} {2019})},\ \Eprint
  {http://arxiv.org/abs/1907.04833} {arXiv:1907.04833 [astro-ph.IM]}
  \BibitemShut {NoStop}%
\bibitem [{\citenamefont {Punturo}\ \emph {et~al.}(2010)\citenamefont {Punturo}
  \emph {et~al.}}]{Punturo:2010zz}%
  \BibitemOpen
  \bibfield  {author} {\bibinfo {author} {\bibfnamefont {M.}~\bibnamefont
  {Punturo}} \emph {et~al.},\ }\bibfield  {booktitle} {\emph {\bibinfo
  {booktitle} {{Proceedings, 14th Workshop on Gravitational wave data analysis
  (GWDAW-14): Rome, Italy, January 26-29, 2010}}},\ }\href {\doibase
  10.1088/0264-9381/27/19/194002} {\bibfield  {journal} {\bibinfo  {journal}
  {Class. Quant. Grav.}\ }\textbf {\bibinfo {volume} {27}},\ \bibinfo {pages}
  {194002} (\bibinfo {year} {2010})}\BibitemShut {NoStop}%
\bibitem [{\citenamefont {Carr}\ \emph {et~al.}(2010)\citenamefont {Carr},
  \citenamefont {Kohri}, \citenamefont {Sendouda},\ and\ \citenamefont
  {Yokoyama}}]{Carr:2009jm}%
  \BibitemOpen
  \bibfield  {author} {\bibinfo {author} {\bibfnamefont {B.~J.}\ \bibnamefont
  {Carr}}, \bibinfo {author} {\bibfnamefont {K.}~\bibnamefont {Kohri}},
  \bibinfo {author} {\bibfnamefont {Y.}~\bibnamefont {Sendouda}}, \ and\
  \bibinfo {author} {\bibfnamefont {J.}~\bibnamefont {Yokoyama}},\ }\href
  {\doibase 10.1103/PhysRevD.81.104019} {\bibfield  {journal} {\bibinfo
  {journal} {Phys. Rev.}\ }\textbf {\bibinfo {volume} {D81}},\ \bibinfo {pages}
  {104019} (\bibinfo {year} {2010})},\ \Eprint {http://arxiv.org/abs/0912.5297}
  {arXiv:0912.5297 [astro-ph.CO]} \BibitemShut {NoStop}%
\bibitem [{\citenamefont {DeRocco}\ and\ \citenamefont
  {Graham}(2019)}]{DeRocco:2019fjq}%
  \BibitemOpen
  \bibfield  {author} {\bibinfo {author} {\bibfnamefont {W.}~\bibnamefont
  {DeRocco}}\ and\ \bibinfo {author} {\bibfnamefont {P.~W.}\ \bibnamefont
  {Graham}},\ }\href {\doibase 10.1103/PhysRevLett.123.251102} {\bibfield
  {journal} {\bibinfo  {journal} {Phys. Rev. Lett.}\ }\textbf {\bibinfo
  {volume} {123}},\ \bibinfo {pages} {251102} (\bibinfo {year} {2019})},\
  \Eprint {http://arxiv.org/abs/1906.07740} {arXiv:1906.07740 [astro-ph.CO]}
  \BibitemShut {NoStop}%
\bibitem [{\citenamefont {Laha}(2019)}]{Laha:2019ssq}%
  \BibitemOpen
  \bibfield  {author} {\bibinfo {author} {\bibfnamefont {R.}~\bibnamefont
  {Laha}},\ }\href {\doibase 10.1103/PhysRevLett.123.251101} {\bibfield
  {journal} {\bibinfo  {journal} {Phys. Rev. Lett.}\ }\textbf {\bibinfo
  {volume} {123}},\ \bibinfo {pages} {251101} (\bibinfo {year} {2019})},\
  \Eprint {http://arxiv.org/abs/1906.09994} {arXiv:1906.09994 [astro-ph.HE]}
  \BibitemShut {NoStop}%
\bibitem [{\citenamefont {Dasgupta}\ \emph {et~al.}(2019)\citenamefont
  {Dasgupta}, \citenamefont {Laha},\ and\ \citenamefont
  {Ray}}]{Dasgupta:2019cae}%
  \BibitemOpen
  \bibfield  {author} {\bibinfo {author} {\bibfnamefont {B.}~\bibnamefont
  {Dasgupta}}, \bibinfo {author} {\bibfnamefont {R.}~\bibnamefont {Laha}}, \
  and\ \bibinfo {author} {\bibfnamefont {A.}~\bibnamefont {Ray}},\ }\href@noop
  {} {\  (\bibinfo {year} {2019})},\ \Eprint {http://arxiv.org/abs/1912.01014}
  {arXiv:1912.01014 [hep-ph]} \BibitemShut {NoStop}%
\bibitem [{\citenamefont {Niikura}\ \emph
  {et~al.}(2019{\natexlab{a}})\citenamefont {Niikura} \emph
  {et~al.}}]{Niikura:2017zjd}%
  \BibitemOpen
  \bibfield  {author} {\bibinfo {author} {\bibfnamefont {H.}~\bibnamefont
  {Niikura}} \emph {et~al.},\ }\href {\doibase 10.1038/s41550-019-0723-1}
  {\bibfield  {journal} {\bibinfo  {journal} {Nat. Astron.}\ }\textbf {\bibinfo
  {volume} {3}},\ \bibinfo {pages} {524} (\bibinfo {year}
  {2019}{\natexlab{a}})},\ \Eprint {http://arxiv.org/abs/1701.02151}
  {arXiv:1701.02151 [astro-ph.CO]} \BibitemShut {NoStop}%
\bibitem [{\citenamefont {Griest}\ \emph {et~al.}(2013)\citenamefont {Griest},
  \citenamefont {Cieplak},\ and\ \citenamefont {Lehner}}]{Griest:2013esa}%
  \BibitemOpen
  \bibfield  {author} {\bibinfo {author} {\bibfnamefont {K.}~\bibnamefont
  {Griest}}, \bibinfo {author} {\bibfnamefont {A.~M.}\ \bibnamefont {Cieplak}},
  \ and\ \bibinfo {author} {\bibfnamefont {M.~J.}\ \bibnamefont {Lehner}},\
  }\href {\doibase 10.1103/PhysRevLett.111.181302} {\bibfield  {journal}
  {\bibinfo  {journal} {Phys. Rev. Lett.}\ }\textbf {\bibinfo {volume} {111}},\
  \bibinfo {pages} {181302} (\bibinfo {year} {2013})}\BibitemShut {NoStop}%
\bibitem [{\citenamefont {Niikura}\ \emph
  {et~al.}(2019{\natexlab{b}})\citenamefont {Niikura}, \citenamefont {Takada},
  \citenamefont {Yokoyama}, \citenamefont {Sumi},\ and\ \citenamefont
  {Masaki}}]{Niikura:2019kqi}%
  \BibitemOpen
  \bibfield  {author} {\bibinfo {author} {\bibfnamefont {H.}~\bibnamefont
  {Niikura}}, \bibinfo {author} {\bibfnamefont {M.}~\bibnamefont {Takada}},
  \bibinfo {author} {\bibfnamefont {S.}~\bibnamefont {Yokoyama}}, \bibinfo
  {author} {\bibfnamefont {T.}~\bibnamefont {Sumi}}, \ and\ \bibinfo {author}
  {\bibfnamefont {S.}~\bibnamefont {Masaki}},\ }\href {\doibase
  10.1103/PhysRevD.99.083503} {\bibfield  {journal} {\bibinfo  {journal} {Phys.
  Rev.}\ }\textbf {\bibinfo {volume} {D99}},\ \bibinfo {pages} {083503}
  (\bibinfo {year} {2019}{\natexlab{b}})},\ \Eprint
  {http://arxiv.org/abs/1901.07120} {arXiv:1901.07120 [astro-ph.CO]}
  \BibitemShut {NoStop}%
\bibitem [{\citenamefont {Allsman}\ \emph {et~al.}(2001)\citenamefont {Allsman}
  \emph {et~al.}}]{Allsman:2000kg}%
  \BibitemOpen
  \bibfield  {author} {\bibinfo {author} {\bibfnamefont {R.~A.}\ \bibnamefont
  {Allsman}} \emph {et~al.} (\bibinfo {collaboration} {Macho}),\ }\href
  {\doibase 10.1086/319636} {\bibfield  {journal} {\bibinfo  {journal}
  {Astrophys. J. Lett.}\ }\textbf {\bibinfo {volume} {550}},\ \bibinfo {pages}
  {L169} (\bibinfo {year} {2001})},\ \Eprint
  {http://arxiv.org/abs/astro-ph/0011506} {arXiv:astro-ph/0011506 [astro-ph]}
  \BibitemShut {NoStop}%
\bibitem [{\citenamefont {Tisserand}\ \emph {et~al.}(2007)\citenamefont
  {Tisserand} \emph {et~al.}}]{Tisserand:2006zx}%
  \BibitemOpen
  \bibfield  {author} {\bibinfo {author} {\bibfnamefont {P.}~\bibnamefont
  {Tisserand}} \emph {et~al.} (\bibinfo {collaboration} {EROS-2}),\ }\href
  {\doibase 10.1051/0004-6361:20066017} {\bibfield  {journal} {\bibinfo
  {journal} {Astron. Astrophys.}\ }\textbf {\bibinfo {volume} {469}},\ \bibinfo
  {pages} {387} (\bibinfo {year} {2007})},\ \Eprint
  {http://arxiv.org/abs/astro-ph/0607207} {arXiv:astro-ph/0607207 [astro-ph]}
  \BibitemShut {NoStop}%
\bibitem [{\citenamefont {Zumalacarregui}\ and\ \citenamefont
  {Seljak}(2018)}]{Zumalacarregui:2017qqd}%
  \BibitemOpen
  \bibfield  {author} {\bibinfo {author} {\bibfnamefont {M.}~\bibnamefont
  {Zumalacarregui}}\ and\ \bibinfo {author} {\bibfnamefont {U.}~\bibnamefont
  {Seljak}},\ }\href {\doibase 10.1103/PhysRevLett.121.141101} {\bibfield
  {journal} {\bibinfo  {journal} {Phys. Rev. Lett.}\ }\textbf {\bibinfo
  {volume} {121}},\ \bibinfo {pages} {141101} (\bibinfo {year} {2018})},\
  \Eprint {http://arxiv.org/abs/1712.02240} {arXiv:1712.02240 [astro-ph.CO]}
  \BibitemShut {NoStop}%
\bibitem [{\citenamefont {Murgia}\ \emph {et~al.}(2019)\citenamefont {Murgia},
  \citenamefont {Scelfo}, \citenamefont {Viel},\ and\ \citenamefont
  {Raccanelli}}]{Murgia:2019duy}%
  \BibitemOpen
  \bibfield  {author} {\bibinfo {author} {\bibfnamefont {R.}~\bibnamefont
  {Murgia}}, \bibinfo {author} {\bibfnamefont {G.}~\bibnamefont {Scelfo}},
  \bibinfo {author} {\bibfnamefont {M.}~\bibnamefont {Viel}}, \ and\ \bibinfo
  {author} {\bibfnamefont {A.}~\bibnamefont {Raccanelli}},\ }\href {\doibase
  10.1103/PhysRevLett.123.071102} {\bibfield  {journal} {\bibinfo  {journal}
  {Phys. Rev. Lett.}\ }\textbf {\bibinfo {volume} {123}},\ \bibinfo {pages}
  {071102} (\bibinfo {year} {2019})},\ \Eprint
  {http://arxiv.org/abs/1903.10509} {arXiv:1903.10509 [astro-ph.CO]}
  \BibitemShut {NoStop}%
\bibitem [{\citenamefont {Poulin}\ \emph {et~al.}(2017)\citenamefont {Poulin},
  \citenamefont {Serpico}, \citenamefont {Calore}, \citenamefont {Clesse},\
  and\ \citenamefont {Kohri}}]{Poulin:2017bwe}%
  \BibitemOpen
  \bibfield  {author} {\bibinfo {author} {\bibfnamefont {V.}~\bibnamefont
  {Poulin}}, \bibinfo {author} {\bibfnamefont {P.~D.}\ \bibnamefont {Serpico}},
  \bibinfo {author} {\bibfnamefont {F.}~\bibnamefont {Calore}}, \bibinfo
  {author} {\bibfnamefont {S.}~\bibnamefont {Clesse}}, \ and\ \bibinfo {author}
  {\bibfnamefont {K.}~\bibnamefont {Kohri}},\ }\href {\doibase
  10.1103/PhysRevD.96.083524} {\bibfield  {journal} {\bibinfo  {journal} {Phys.
  Rev.}\ }\textbf {\bibinfo {volume} {D96}},\ \bibinfo {pages} {083524}
  (\bibinfo {year} {2017})},\ \Eprint {http://arxiv.org/abs/1707.04206}
  {arXiv:1707.04206 [astro-ph.CO]} \BibitemShut {NoStop}%
\bibitem [{\citenamefont {Vaskonen}\ and\ \citenamefont
  {Veermäe}(2020)}]{Vaskonen:2019jpv}%
  \BibitemOpen
  \bibfield  {author} {\bibinfo {author} {\bibfnamefont {V.}~\bibnamefont
  {Vaskonen}}\ and\ \bibinfo {author} {\bibfnamefont {H.}~\bibnamefont
  {Veermäe}},\ }\href {\doibase 10.1103/PhysRevD.101.043015} {\bibfield
  {journal} {\bibinfo  {journal} {Phys. Rev.}\ }\textbf {\bibinfo {volume}
  {D101}},\ \bibinfo {pages} {043015} (\bibinfo {year} {2020})},\ \Eprint
  {http://arxiv.org/abs/1908.09752} {arXiv:1908.09752 [astro-ph.CO]}
  \BibitemShut {NoStop}%
\bibitem [{\citenamefont {Kamionkowski}\ \emph {et~al.}(1994)\citenamefont
  {Kamionkowski}, \citenamefont {Kosowsky},\ and\ \citenamefont
  {Turner}}]{Kamionkowski:1993fg}%
  \BibitemOpen
  \bibfield  {author} {\bibinfo {author} {\bibfnamefont {M.}~\bibnamefont
  {Kamionkowski}}, \bibinfo {author} {\bibfnamefont {A.}~\bibnamefont
  {Kosowsky}}, \ and\ \bibinfo {author} {\bibfnamefont {M.~S.}\ \bibnamefont
  {Turner}},\ }\href {\doibase 10.1103/PhysRevD.49.2837} {\bibfield  {journal}
  {\bibinfo  {journal} {Phys. Rev.}\ }\textbf {\bibinfo {volume} {D49}},\
  \bibinfo {pages} {2837} (\bibinfo {year} {1994})},\ \Eprint
  {http://arxiv.org/abs/astro-ph/9310044} {arXiv:astro-ph/9310044 [astro-ph]}
  \BibitemShut {NoStop}%
\bibitem [{\citenamefont {Huber}\ and\ \citenamefont
  {Konstandin}(2008)}]{Huber:2008hg}%
  \BibitemOpen
  \bibfield  {author} {\bibinfo {author} {\bibfnamefont {S.~J.}\ \bibnamefont
  {Huber}}\ and\ \bibinfo {author} {\bibfnamefont {T.}~\bibnamefont
  {Konstandin}},\ }\href {\doibase 10.1088/1475-7516/2008/09/022} {\bibfield
  {journal} {\bibinfo  {journal} {JCAP}\ }\textbf {\bibinfo {volume} {0809}},\
  \bibinfo {pages} {022} (\bibinfo {year} {2008})},\ \Eprint
  {http://arxiv.org/abs/0806.1828} {arXiv:0806.1828 [hep-ph]} \BibitemShut
  {NoStop}%
\bibitem [{\citenamefont {Guth}(1982)}]{Guth:1982cv}%
  \BibitemOpen
  \bibfield  {author} {\bibinfo {author} {\bibfnamefont {A.~H.}\ \bibnamefont
  {Guth}},\ }in\ \href@noop {} {\emph {\bibinfo {booktitle} {{Nuffield Workshop
  on the Very Early Universe}}}}\ (\bibinfo {year} {1982})\BibitemShut
  {NoStop}%
\bibitem [{\citenamefont {Ellis}\ \emph
  {et~al.}(2019{\natexlab{b}})\citenamefont {Ellis}, \citenamefont {Lewicki},\
  and\ \citenamefont {No}}]{Ellis:2018mja}%
  \BibitemOpen
  \bibfield  {author} {\bibinfo {author} {\bibfnamefont {J.}~\bibnamefont
  {Ellis}}, \bibinfo {author} {\bibfnamefont {M.}~\bibnamefont {Lewicki}}, \
  and\ \bibinfo {author} {\bibfnamefont {J.~M.}\ \bibnamefont {No}},\ }\href
  {\doibase 10.1088/1475-7516/2019/04/003} {\bibfield  {journal} {\bibinfo
  {journal} {JCAP}\ }\textbf {\bibinfo {volume} {04}},\ \bibinfo {pages} {003}
  (\bibinfo {year} {2019}{\natexlab{b}})},\ \Eprint
  {http://arxiv.org/abs/1809.08242} {arXiv:1809.08242 [hep-ph]} \BibitemShut
  {NoStop}%
\bibitem [{\citenamefont {Peccei}\ and\ \citenamefont
  {Quinn}(1977)}]{Peccei:1977ur}%
  \BibitemOpen
  \bibfield  {author} {\bibinfo {author} {\bibfnamefont {R.~D.}\ \bibnamefont
  {Peccei}}\ and\ \bibinfo {author} {\bibfnamefont {H.~R.}\ \bibnamefont
  {Quinn}},\ }\href {\doibase 10.1103/PhysRevD.16.1791} {\bibfield  {journal}
  {\bibinfo  {journal} {Phys. Rev.}\ }\textbf {\bibinfo {volume} {D16}},\
  \bibinfo {pages} {1791} (\bibinfo {year} {1977})}\BibitemShut {NoStop}%
\bibitem [{\citenamefont {Chacko}\ \emph {et~al.}(2004)\citenamefont {Chacko},
  \citenamefont {Hall},\ and\ \citenamefont {Nomura}}]{Chacko:2004ky}%
  \BibitemOpen
  \bibfield  {author} {\bibinfo {author} {\bibfnamefont {Z.}~\bibnamefont
  {Chacko}}, \bibinfo {author} {\bibfnamefont {L.~J.}\ \bibnamefont {Hall}}, \
  and\ \bibinfo {author} {\bibfnamefont {Y.}~\bibnamefont {Nomura}},\ }\href
  {\doibase 10.1088/1475-7516/2004/10/011} {\bibfield  {journal} {\bibinfo
  {journal} {JCAP}\ }\textbf {\bibinfo {volume} {0410}},\ \bibinfo {pages}
  {011} (\bibinfo {year} {2004})},\ \Eprint
  {http://arxiv.org/abs/astro-ph/0405596} {arXiv:astro-ph/0405596 [astro-ph]}
  \BibitemShut {NoStop}%
\bibitem [{\citenamefont {Romero}\ \emph {et~al.}(2021)\citenamefont {Romero},
  \citenamefont {Martinovic}, \citenamefont {Callister}, \citenamefont {Guo},
  \citenamefont {Martínez}, \citenamefont {Sakellariadou}, \citenamefont
  {Yang},\ and\ \citenamefont {Zhao}}]{Romero:2021kby}%
  \BibitemOpen
  \bibfield  {author} {\bibinfo {author} {\bibfnamefont {A.}~\bibnamefont
  {Romero}}, \bibinfo {author} {\bibfnamefont {K.}~\bibnamefont {Martinovic}},
  \bibinfo {author} {\bibfnamefont {T.~A.}\ \bibnamefont {Callister}}, \bibinfo
  {author} {\bibfnamefont {H.-K.}\ \bibnamefont {Guo}}, \bibinfo {author}
  {\bibfnamefont {M.}~\bibnamefont {Martínez}}, \bibinfo {author}
  {\bibfnamefont {M.}~\bibnamefont {Sakellariadou}}, \bibinfo {author}
  {\bibfnamefont {F.-W.}\ \bibnamefont {Yang}}, \ and\ \bibinfo {author}
  {\bibfnamefont {Y.}~\bibnamefont {Zhao}},\ }\href {\doibase
  10.1103/PhysRevLett.126.151301} {\bibfield  {journal} {\bibinfo  {journal}
  {Phys. Rev. Lett.}\ }\textbf {\bibinfo {volume} {126}},\ \bibinfo {pages}
  {151301} (\bibinfo {year} {2021})},\ \Eprint
  {http://arxiv.org/abs/2102.01714} {arXiv:2102.01714 [hep-ph]} \BibitemShut
  {NoStop}%
\bibitem [{\citenamefont {Arzoumanian}\ \emph {et~al.}(2020)\citenamefont
  {Arzoumanian} \emph {et~al.}}]{Arzoumanian:2020vkk}%
  \BibitemOpen
  \bibfield  {author} {\bibinfo {author} {\bibfnamefont {Z.}~\bibnamefont
  {Arzoumanian}} \emph {et~al.} (\bibinfo {collaboration} {NANOGrav}),\ }\href
  {\doibase 10.3847/2041-8213/abd401} {\bibfield  {journal} {\bibinfo
  {journal} {Astrophys. J. Lett.}\ }\textbf {\bibinfo {volume} {905}},\
  \bibinfo {pages} {L34} (\bibinfo {year} {2020})},\ \Eprint
  {http://arxiv.org/abs/2009.04496} {arXiv:2009.04496 [astro-ph.HE]}
  \BibitemShut {NoStop}%
\bibitem [{\citenamefont {Bian}\ \emph
  {et~al.}(2021{\natexlab{b}})\citenamefont {Bian}, \citenamefont {Cai},
  \citenamefont {Liu}, \citenamefont {Yang},\ and\ \citenamefont
  {Zhou}}]{Bian:2021lmz}%
  \BibitemOpen
  \bibfield  {author} {\bibinfo {author} {\bibfnamefont {L.}~\bibnamefont
  {Bian}}, \bibinfo {author} {\bibfnamefont {R.-G.}\ \bibnamefont {Cai}},
  \bibinfo {author} {\bibfnamefont {J.}~\bibnamefont {Liu}}, \bibinfo {author}
  {\bibfnamefont {X.-Y.}\ \bibnamefont {Yang}}, \ and\ \bibinfo {author}
  {\bibfnamefont {R.}~\bibnamefont {Zhou}},\ }\href {\doibase
  10.1103/PhysRevD.103.L081301} {\bibfield  {journal} {\bibinfo  {journal}
  {Phys. Rev.}\ }\textbf {\bibinfo {volume} {D103}},\ \bibinfo {pages}
  {L081301} (\bibinfo {year} {2021}{\natexlab{b}})},\ \Eprint
  {http://arxiv.org/abs/2009.13893} {arXiv:2009.13893 [astro-ph.CO]}
  \BibitemShut {NoStop}%
\bibitem [{\citenamefont {Deng}(2021)}]{Deng:2021ezy}%
  \BibitemOpen
  \bibfield  {author} {\bibinfo {author} {\bibfnamefont {H.}~\bibnamefont
  {Deng}},\ }\href {\doibase 10.1088/1475-7516/2021/04/058} {\bibfield
  {journal} {\bibinfo  {journal} {JCAP}\ }\textbf {\bibinfo {volume} {2104}},\
  \bibinfo {pages} {058} (\bibinfo {year} {2021})},\ \Eprint
  {http://arxiv.org/abs/2101.11098} {arXiv:2101.11098 [astro-ph.CO]}
  \BibitemShut {NoStop}%
\bibitem [{\citenamefont {Kosowsky}\ and\ \citenamefont
  {Turner}(1993)}]{Kosowsky:1992vn}%
  \BibitemOpen
  \bibfield  {author} {\bibinfo {author} {\bibfnamefont {A.}~\bibnamefont
  {Kosowsky}}\ and\ \bibinfo {author} {\bibfnamefont {M.~S.}\ \bibnamefont
  {Turner}},\ }\href {\doibase 10.1103/PhysRevD.47.4372} {\bibfield  {journal}
  {\bibinfo  {journal} {Phys. Rev.}\ }\textbf {\bibinfo {volume} {D47}},\
  \bibinfo {pages} {4372} (\bibinfo {year} {1993})},\ \Eprint
  {http://arxiv.org/abs/astro-ph/9211004} {arXiv:astro-ph/9211004 [astro-ph]}
  \BibitemShut {NoStop}%
\bibitem [{\citenamefont {Weir}(2016)}]{Weir:2016tov}%
  \BibitemOpen
  \bibfield  {author} {\bibinfo {author} {\bibfnamefont {D.~J.}\ \bibnamefont
  {Weir}},\ }\href {\doibase 10.1103/PhysRevD.93.124037} {\bibfield  {journal}
  {\bibinfo  {journal} {Phys. Rev.}\ }\textbf {\bibinfo {volume} {D93}},\
  \bibinfo {pages} {124037} (\bibinfo {year} {2016})},\ \Eprint
  {http://arxiv.org/abs/1604.08429} {arXiv:1604.08429 [astro-ph.CO]}
  \BibitemShut {NoStop}%
\bibitem [{\citenamefont {Konstandin}(2018)}]{Konstandin:2017sat}%
  \BibitemOpen
  \bibfield  {author} {\bibinfo {author} {\bibfnamefont {T.}~\bibnamefont
  {Konstandin}},\ }\href {\doibase 10.1088/1475-7516/2018/03/047} {\bibfield
  {journal} {\bibinfo  {journal} {JCAP}\ }\textbf {\bibinfo {volume} {03}},\
  \bibinfo {pages} {047} (\bibinfo {year} {2018})},\ \Eprint
  {http://arxiv.org/abs/1712.06869} {arXiv:1712.06869 [astro-ph.CO]}
  \BibitemShut {NoStop}%
\bibitem [{\citenamefont {Jinno}\ and\ \citenamefont
  {Takimoto}(2019)}]{Jinno:2017fby}%
  \BibitemOpen
  \bibfield  {author} {\bibinfo {author} {\bibfnamefont {R.}~\bibnamefont
  {Jinno}}\ and\ \bibinfo {author} {\bibfnamefont {M.}~\bibnamefont
  {Takimoto}},\ }\href {\doibase 10.1088/1475-7516/2019/01/060} {\bibfield
  {journal} {\bibinfo  {journal} {JCAP}\ }\textbf {\bibinfo {volume} {01}},\
  \bibinfo {pages} {060} (\bibinfo {year} {2019})},\ \Eprint
  {http://arxiv.org/abs/1707.03111} {arXiv:1707.03111 [hep-ph]} \BibitemShut
  {NoStop}%
\bibitem [{\citenamefont {Hindmarsh}(2018)}]{Hindmarsh:2016lnk}%
  \BibitemOpen
  \bibfield  {author} {\bibinfo {author} {\bibfnamefont {M.}~\bibnamefont
  {Hindmarsh}},\ }\href {\doibase 10.1103/PhysRevLett.120.071301} {\bibfield
  {journal} {\bibinfo  {journal} {Phys. Rev. Lett.}\ }\textbf {\bibinfo
  {volume} {120}},\ \bibinfo {pages} {071301} (\bibinfo {year} {2018})},\
  \Eprint {http://arxiv.org/abs/1608.04735} {arXiv:1608.04735 [astro-ph.CO]}
  \BibitemShut {NoStop}%
\bibitem [{\citenamefont {Hindmarsh}\ and\ \citenamefont
  {Hijazi}(2019)}]{Hindmarsh:2019phv}%
  \BibitemOpen
  \bibfield  {author} {\bibinfo {author} {\bibfnamefont {M.}~\bibnamefont
  {Hindmarsh}}\ and\ \bibinfo {author} {\bibfnamefont {M.}~\bibnamefont
  {Hijazi}},\ }\href {\doibase 10.1088/1475-7516/2019/12/062} {\bibfield
  {journal} {\bibinfo  {journal} {JCAP}\ }\textbf {\bibinfo {volume} {1912}},\
  \bibinfo {pages} {062} (\bibinfo {year} {2019})},\ \Eprint
  {http://arxiv.org/abs/1909.10040} {arXiv:1909.10040 [astro-ph.CO]}
  \BibitemShut {NoStop}%
\bibitem [{\citenamefont {Hindmarsh}\ \emph {et~al.}(2014)\citenamefont
  {Hindmarsh}, \citenamefont {Huber}, \citenamefont {Rummukainen},\ and\
  \citenamefont {Weir}}]{Hindmarsh:2013xza}%
  \BibitemOpen
  \bibfield  {author} {\bibinfo {author} {\bibfnamefont {M.}~\bibnamefont
  {Hindmarsh}}, \bibinfo {author} {\bibfnamefont {S.~J.}\ \bibnamefont
  {Huber}}, \bibinfo {author} {\bibfnamefont {K.}~\bibnamefont {Rummukainen}},
  \ and\ \bibinfo {author} {\bibfnamefont {D.~J.}\ \bibnamefont {Weir}},\
  }\href {\doibase 10.1103/PhysRevLett.112.041301} {\bibfield  {journal}
  {\bibinfo  {journal} {Phys. Rev. Lett.}\ }\textbf {\bibinfo {volume} {112}},\
  \bibinfo {pages} {041301} (\bibinfo {year} {2014})},\ \Eprint
  {http://arxiv.org/abs/1304.2433} {arXiv:1304.2433 [hep-ph]} \BibitemShut
  {NoStop}%
\bibitem [{\citenamefont {Hindmarsh}\ \emph {et~al.}(2015)\citenamefont
  {Hindmarsh}, \citenamefont {Huber}, \citenamefont {Rummukainen},\ and\
  \citenamefont {Weir}}]{Hindmarsh:2015qta}%
  \BibitemOpen
  \bibfield  {author} {\bibinfo {author} {\bibfnamefont {M.}~\bibnamefont
  {Hindmarsh}}, \bibinfo {author} {\bibfnamefont {S.~J.}\ \bibnamefont
  {Huber}}, \bibinfo {author} {\bibfnamefont {K.}~\bibnamefont {Rummukainen}},
  \ and\ \bibinfo {author} {\bibfnamefont {D.~J.}\ \bibnamefont {Weir}},\
  }\href {\doibase 10.1103/PhysRevD.92.123009} {\bibfield  {journal} {\bibinfo
  {journal} {Phys. Rev.}\ }\textbf {\bibinfo {volume} {D92}},\ \bibinfo {pages}
  {123009} (\bibinfo {year} {2015})},\ \Eprint
  {http://arxiv.org/abs/1504.03291} {arXiv:1504.03291 [astro-ph.CO]}
  \BibitemShut {NoStop}%
\bibitem [{\citenamefont {Cutting}\ \emph {et~al.}(2018)\citenamefont
  {Cutting}, \citenamefont {Hindmarsh},\ and\ \citenamefont
  {Weir}}]{Cutting:2018tjt}%
  \BibitemOpen
  \bibfield  {author} {\bibinfo {author} {\bibfnamefont {D.}~\bibnamefont
  {Cutting}}, \bibinfo {author} {\bibfnamefont {M.}~\bibnamefont {Hindmarsh}},
  \ and\ \bibinfo {author} {\bibfnamefont {D.~J.}\ \bibnamefont {Weir}},\
  }\href {\doibase 10.1103/PhysRevD.97.123513} {\bibfield  {journal} {\bibinfo
  {journal} {Phys. Rev.}\ }\textbf {\bibinfo {volume} {D97}},\ \bibinfo {pages}
  {123513} (\bibinfo {year} {2018})},\ \Eprint
  {http://arxiv.org/abs/1802.05712} {arXiv:1802.05712 [astro-ph.CO]}
  \BibitemShut {NoStop}%
\bibitem [{\citenamefont {Cutting}\ \emph {et~al.}(2021)\citenamefont
  {Cutting}, \citenamefont {Escartin}, \citenamefont {Hindmarsh},\ and\
  \citenamefont {Weir}}]{Cutting:2020nla}%
  \BibitemOpen
  \bibfield  {author} {\bibinfo {author} {\bibfnamefont {D.}~\bibnamefont
  {Cutting}}, \bibinfo {author} {\bibfnamefont {E.~G.}\ \bibnamefont
  {Escartin}}, \bibinfo {author} {\bibfnamefont {M.}~\bibnamefont {Hindmarsh}},
  \ and\ \bibinfo {author} {\bibfnamefont {D.~J.}\ \bibnamefont {Weir}},\
  }\href {\doibase 10.1103/PhysRevD.103.023531} {\bibfield  {journal} {\bibinfo
   {journal} {Phys. Rev.}\ }\textbf {\bibinfo {volume} {D103}},\ \bibinfo
  {pages} {023531} (\bibinfo {year} {2021})},\ \Eprint
  {http://arxiv.org/abs/2005.13537} {arXiv:2005.13537 [astro-ph.CO]}
  \BibitemShut {NoStop}%
\bibitem [{\citenamefont {Di}\ \emph {et~al.}(2020)\citenamefont {Di},
  \citenamefont {Wang}, \citenamefont {Zhou}, \citenamefont {Bian},
  \citenamefont {Cai},\ and\ \citenamefont {Liu}}]{Di:2020nny}%
  \BibitemOpen
  \bibfield  {author} {\bibinfo {author} {\bibfnamefont {Y.}~\bibnamefont
  {Di}}, \bibinfo {author} {\bibfnamefont {J.}~\bibnamefont {Wang}}, \bibinfo
  {author} {\bibfnamefont {R.}~\bibnamefont {Zhou}}, \bibinfo {author}
  {\bibfnamefont {L.}~\bibnamefont {Bian}}, \bibinfo {author} {\bibfnamefont
  {R.-G.}\ \bibnamefont {Cai}}, \ and\ \bibinfo {author} {\bibfnamefont
  {J.}~\bibnamefont {Liu}},\ }\href@noop {} {\  (\bibinfo {year} {2020})},\
  \Eprint {http://arxiv.org/abs/2012.15625} {arXiv:2012.15625 [astro-ph.CO]}
  \BibitemShut {NoStop}%
\bibitem [{\citenamefont {Caprini}\ \emph {et~al.}(2016)\citenamefont {Caprini}
  \emph {et~al.}}]{Caprini:2015zlo}%
  \BibitemOpen
  \bibfield  {author} {\bibinfo {author} {\bibfnamefont {C.}~\bibnamefont
  {Caprini}} \emph {et~al.},\ }\href {\doibase 10.1088/1475-7516/2016/04/001}
  {\bibfield  {journal} {\bibinfo  {journal} {JCAP}\ }\textbf {\bibinfo
  {volume} {1604}},\ \bibinfo {pages} {001} (\bibinfo {year} {2016})},\ \Eprint
  {http://arxiv.org/abs/1512.06239} {arXiv:1512.06239 [astro-ph.CO]}
  \BibitemShut {NoStop}%
\bibitem [{\citenamefont {Cai}\ \emph {et~al.}(2017{\natexlab{b}})\citenamefont
  {Cai}, \citenamefont {Sasaki},\ and\ \citenamefont {Wang}}]{Cai:2017tmh}%
  \BibitemOpen
  \bibfield  {author} {\bibinfo {author} {\bibfnamefont {R.-G.}\ \bibnamefont
  {Cai}}, \bibinfo {author} {\bibfnamefont {M.}~\bibnamefont {Sasaki}}, \ and\
  \bibinfo {author} {\bibfnamefont {S.-J.}\ \bibnamefont {Wang}},\ }\href
  {\doibase 10.1088/1475-7516/2017/08/004} {\bibfield  {journal} {\bibinfo
  {journal} {JCAP}\ }\textbf {\bibinfo {volume} {1708}},\ \bibinfo {pages}
  {004} (\bibinfo {year} {2017}{\natexlab{b}})},\ \Eprint
  {http://arxiv.org/abs/1707.03001} {arXiv:1707.03001 [astro-ph.CO]}
  \BibitemShut {NoStop}%
\bibitem [{\citenamefont {Musco}\ and\ \citenamefont
  {Miller}(2013)}]{Musco:2012au}%
  \BibitemOpen
  \bibfield  {author} {\bibinfo {author} {\bibfnamefont {I.}~\bibnamefont
  {Musco}}\ and\ \bibinfo {author} {\bibfnamefont {J.~C.}\ \bibnamefont
  {Miller}},\ }\href {\doibase 10.1088/0264-9381/30/14/145009} {\bibfield
  {journal} {\bibinfo  {journal} {Class. Quant. Grav.}\ }\textbf {\bibinfo
  {volume} {30}},\ \bibinfo {pages} {145009} (\bibinfo {year} {2013})},\
  \Eprint {http://arxiv.org/abs/1201.2379} {arXiv:1201.2379 [gr-qc]}
  \BibitemShut {NoStop}%
\end{thebibliography}%
\end{document}